\documentclass[11pt]{article}
\usepackage{epsfig,psfig,float,amssymb,latexsym}
\usepackage[notref,notcite]{}

\newcommand{\be}{\begin{equation}}
\newcommand{\ee}{\end{equation}}
\newcommand{\ba}{\begin{eqnarray}}
\newcommand{\ea}{\end{eqnarray}}

\newcommand{\nn}{\nonumber}

\newcommand{\rhoslash}{\rho\!\!\!/}

\begin{document}
\hfill{IFIC$-$00$-$1108}

\hfill{FTUV$-$00$-$1108}
\begin{center}
{\Huge{\bf{Chiral approach to the rho meson\\
\vspace{0.4cm}
in nuclear matter}}}
\end{center}
\vspace{.5cm}

\begin{center}
{\huge{D. Cabrera, E. Oset and M.J. Vicente Vacas}}
\end{center}

\begin{center}
{\small{\it Departamento de F\'{\i}sica Te\'orica and IFIC,\\
Centro Mixto Universidad de Valencia-CSIC,\\
Institutos de Investigaci\'on de Paterna, Apdo. correos 22085,\\
46071, Valencia, Spain}}
\end{center}

\vspace{1cm}

\begin{abstract}
{\small{In this work, the properties of the $\rho$ meson at rest in cold
symmetric nuclear matter are studied. We make use of a chiral unitary approach 
to pion-pion scattering in the vector-isovector channel, calculated from the 
lowest order Chiral Perturbation Theory ($\chi PT$) lagrangian including 
explicit resonance fields. Low energy chiral constraints are considered by 
matching our expressions to those of one loop $\chi PT$. To account for the 
medium corrections, the $\rho$ couples to $\pi\pi$ pairs which are properly 
renormalized in the nuclear medium, accounting for both $p-h$ and $\Delta -h$ 
excitations. The terms where the $\rho$ couples directly to the hadrons in the
$p-h$ or $\Delta-h$ excitations are also accounted for. In addition, the $\rho$
is also allowed to couple to $N^{*}(1520)-h$ components.

\vspace{0.5cm}

}}
\end{abstract}

\section{Introduction}
   One of the major goals of nuclear physics is the understanding of the
properties of hadrons at the high baryonic densities that occur in  nuclei or
neutron stars.  However, it is difficult to extract information on the mass or
width  of an in-medium meson from experimental data because in most  of the
decay channels the final particles  undergo  a strong distortion before they
get out from the dense matter and can  reach the detector. 

Electromagnetic decays offer, at least in principle, a cleaner probe  of the
high density regions. Neutral vector mesons, like $\rho^0$,  are specially
interesting due to its large leptonic width. Any changes in the mass or width
of these mesons should be reflected in the invariant  mass distributions of the
leptonic decay products. 

Although there are some problems in the interpretation of data, due to  the low
statistics and the existence of many additional sources, dilepton  spectra,
such as those measured at CERN
\cite{Agakishiev:1995xb,Lenkeit:1999xu,Filimonov:2001fd} and at
lower energies at Bevalac \cite{Porter:1997rc,Ozawa:2001iw}, may indicate either
a  lowering 
of the $\rho$ mass or a large broadening of its width.  In the near future,
experiments at GSI (HADES Collaboration 
\cite{Friese:1999qm,Bratkovskaya:2001mb}) could clarify the situation by  
providing better statistics and mass resolution.

Many theoretical approaches have been pursued to analyse the rho
meson properties in the medium. A good review of the current situation can be
found in ref. \cite{Rapp:2000ej}. Here, we only present a brief description
of the main lines of work. Much of the current interest was stimulated
by Brown and Rho \cite{Brown:1991kk} who by using scale invariance arguments,
obtained an approximate in-medium scaling law predicting the $\rho$ mass
to decrease as a function of density. Also Hatsuda and Lee \cite{Hatsuda:1992ez}
found a linear decrease of the masses as a function of density in a QCD sum
rules calculation, although recent works
\cite{Leupold:1998dg,Klingl:1997kf,Mallik:2001gv} 
cast some doubts on these conclusions and find that predictions from QCD
sum rules are also consistent with larger $\rho$ masses if the width is
broad enough. Several groups have investigated the $\rho$ selfenergy in the
nuclear medium studying  the decay into two pions, the coupling to nucleons
to excite a baryonic resonance, or both. The main effects on the two pion decay
channel are produced by the coupling  to $\Delta$-hole  excitations
\cite{Asakawa:1992ht,Asakawa:1993pq,Chanfray:1993ue,Herrmann:1993za,
Urban:1998eg,Urban:2000im,Broniowski:2001qd}. Whereas the $\rho$ mass is barely
changed, quite a  
large broadening, which provides a considerable strength at low masses, is 
obtained in these calculations. Another strong source of selfenergy is 
the excitation of baryonic resonances, mainly the $N^*(1520)$ considered in
refs. \cite{Rapp:1997fs,Peters:1998va,Post:2001qi}, which further broadens
the $\rho$ meson. Finally, we could mention the systematic coupled channel 
approach to meson-nucleon scattering of ref. \cite{Lutz:1999jn} that
automatically provides the first order, in a density expansion, of the
$\rho$ selfenergy, finding also a large spreading of the strength to states at
low energy.

 In this paper we investigate the rho meson properties in cold nuclear matter
in a  non-perturbative coupled channels chiral model. This approach combines
constraints from chiral symmetry breaking and unitarity, and has proved very
successful in describing mesonic properties in vacuum at low  and intermediate
energies \cite{Oller:1999hw}. The effects of  the nuclear medium on the mesons
have already been analyzed in this framework,  for the scalar isoscalar
$(I=J=0),\;"\sigma"$ channel \cite{Chiang:1998di,Oset:2000ev}. There have been
several attempts to translate  the nuclear matter results into observable
magnitudes, which can be compared  with experimental
results\cite{Rapp:1999fx,VicenteVacas:1999xx}. However, as mentioned before, it
is difficult  to disentangle the meson properties at high baryonic densities
due to the  strong distortion of the decay products. The rho meson studied here
is better suited for such a study because of its large leptonic width.

In section 2 we give a review of the model for meson-meson scattering in
vacuum, presenting results for the $\rho$ channel. Nuclear medium corrections
are studied in section 3, with special attention to requests of gauge
invariance. In section 4 we describe how the coupling to
$N^*(1520) -h$ components can be incorporated in the calculation. Our results
are presented in section 5. Finally we summarize in section 6.

\section{Meson-meson scattering in a chiral unitary approach}
In this work we study the $\rho$ propagation properties by obtaining the 
$\pi \pi \to \pi \pi$  scattering amplitude in the $(I,J)=(1,1)$ channel. We 
first give a brief description of the model for meson-meson scattering in 
vacuum, and then discuss modifications arising in the presence of 
cold symmetric nuclear matter. The model for meson-meson scattering in vacuum is
explained in detail in \cite{Oller:2001ug}. Following that work, we stay in the frame of
a coupled channel chiral unitary approach starting from the tree level graphs 
from the lowest order $\chi PT$ lagrangians \cite{gasser1} including explicit 
resonance fields \cite{gasser2}. The model successfully describes $\pi\pi$ 
P-wave phase shifts and $\pi$, $K$ electromagnetic vector form factors up to 
$\sqrt{s}\lesssim 1.2$ GeV. Low energy chiral constraints are satisfied by a 
matching of expressions with one-loop $\chi PT$. Unitarity is imposed following 
the $N/D$ method.

We start from the $(I=1)$ $\pi\pi$, $K \bar{K}$ states in the isospin basis,
using the unitarity normalization \cite{Oller:1997ti}:
\ba
\label{isospin}
|\pi\pi\rangle&=&\frac{1}{2}|\pi^+\pi^- - \pi^- \pi^+\rangle\nn \\
|K\bar{K}\rangle&=& \frac{1}{\sqrt{2}} |K^+K^- - K^0\bar{K}^0\rangle.
\ea
Tree level amplitudes are collected in a $2\times 2$ $K$ matrix whose elements 
are

\ba
\label{treeK}
K_{11}(s)&=& \frac{1}{3}\frac{p_{1}^2}{f^2} \left[1+
\frac{2\,G_V^2}{f^2} \frac{s}{M_\rho^2-s}\right]\nn \\
K_{12}(s)&=& \frac{\sqrt{2}}{3}\frac{p_{1} \,p_{2}}{f^2}
\left[1+\frac{2\,G_V^2}{f^2} \frac{s}{M_\rho^2-s}\right] \nn \\
K_{21}(s)&=&K_{12}(s) \nn \\
K_{22}(s)&=& \frac{2}{3}\frac{p_{2}^2}{f^2} \left[1+
\frac{2\,G_V^2}{f^2} \frac{s}{M_\rho^2-s}\right]
\ea
with the labels 1 for $K \bar{K}$ and 2 for $\pi\pi$ states. In equation 
(\ref{treeK}) $G_{V}$ is the strength of the pseudoscalar-vector resonance 
vertex, $f$ the pion decay constant in the chiral limit, $s$ the squared 
invariant mass, $M_{\rho}$ the bare mass of the $\rho$ meson and
$p_{i}=\sqrt{s/4-m_{i}^{2}}$. In any of the amplitudes of eq. (\ref{treeK}) the 
first term comes from the $\mathcal{O}(p^2)$ chiral lagrangian, while the 
second term is the contribution of the resonance lagrangian \cite{Oller:2001ug}.

The approach to meson-meson scattering in the isovector channel commented so far 
is essentially equivalent to using gauge vector fields for the $\rho$ meson due
to VMD. We will use the gauge properties of the $\rho$ mesons in order to
evaluate vertex corrections which appear when selfenergy insertions are also
considered.

The final expression of the $T$ matrix is obtained by unitarizing the tree level 
scattering amplitudes in eq. (\ref{treeK}). To this end we follow the N/D 
method, which was adapted to the context of chiral theory in ref. \cite{N/D}. We 
get
\be
\label{T}
T(s)= \left[I+K(s)\cdot G(s) \right]^{-1}\cdot K(s),
\ee
where $G(s)$ is a diagonal matrix given by the loop integral of two meson
propagators. In dimensional regularization its diagonal elements are given by

\be
\label{g(s)}
G_i^D(s)=\frac{1}{16\,\pi^2}\left[-2+d_i+\sigma_{i}(s)\,
\log \frac{\sigma_{i}(s)+1}{\sigma_{i}(s)-1} \right],
\ee
where the subindex $i$ refers to the corresponding two meson state and
$\sigma_{i}(s)=\sqrt{1-4 m_i^2/s}$ with $m_i$ the mass of the particles in the
state $i$.
 
\begin{figure}[ht]
\centerline{\includegraphics[width=0.7\textwidth]{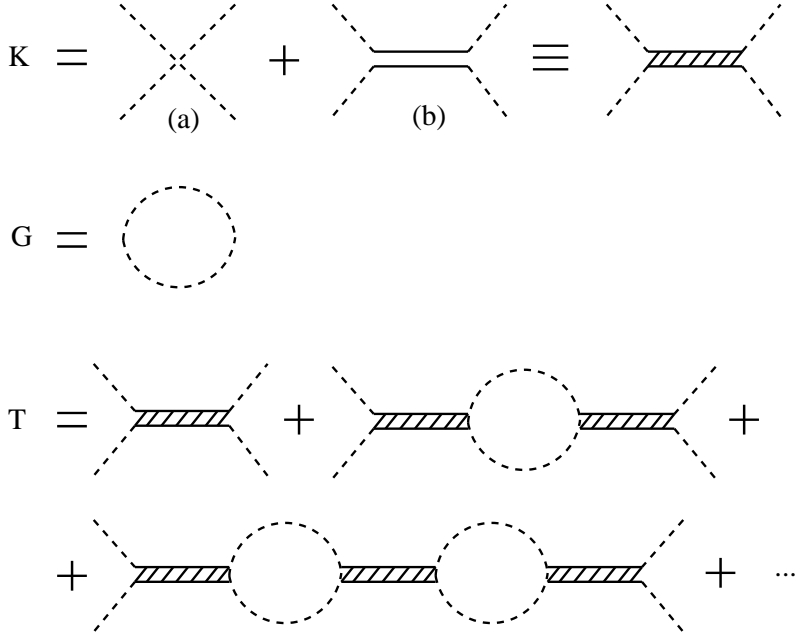}}
\caption{\footnotesize{Diagrammatic representation of the matrices described in 
the text. (a) and (b) correspond to the amplitudes derived from the 
$\mathcal{O}(p^2)$ and resonance chiral lagrangians respectively. The double
solid line filled with short oblique dashes represents the effective scattering
amplitude. Dashed lines are $\pi$, $K$ mesons.}}
\label{diagmatrix}
\end{figure}
In Fig. \ref{diagmatrix} we show a diagrammatic representation of the three 
matrices involved, $K(s)$, $G(s)$ and $T(s)$. The last one represents the 
resummation performed in eq. (\ref{T}) with the unitarization procedure (N/D 
method).

The $d_i$ constants in eq. (\ref{g(s)}) are chosen to obey the low energy chiral 
constraints \cite{Oller:2001ug},

\ba
\label{dconstants}
d_K=\frac{-2\ m_\pi^2}{m_K^2-m_\pi^2}\left(\log\frac{m_\pi^2}
{\mu^2}+\frac{1}{2} \log\frac{m_K^2}{\mu^2}+\frac{1}{2}\right)\nn\\
d_\pi=\frac{m_K^2}{m_K^2-m_\pi^2}\left(\log\frac{m_\pi^2}{\mu^2}+\frac{1}{2}
\log\frac{m_K^2}{\mu^2}+\frac{1}{2}\right),
\ea
and they are obtained by matching the expressions of the form factors 
calculated in this approach with those of one loop $\chi PT$. In eq. 
(\ref{dconstants}) $\mu = 770$ MeV and we have substituted index values
$i=1,2$ by their actual meaning.

The regularization can also be carried out following a cut-off scheme. The loop
functions with a cut-off in the three-momentum of the particles in the loop can
be found in the appendix of ref. \cite{Oller:1999hw}, and up to order
$\frac{m_i}{q_i^{max}}$ they read

\be
\label{g(s)cutoff}
G^C_i(s)=\frac{1}{16\,\pi^2}\left[-2\,
\log \frac{2\,q_i^{max}}{m_i}+\sigma_{i}(s)\,
\log \frac{\sigma_{i}(s)+1}{\sigma_{i}(s)-1} \right],
\ee
where $q_i^{max}$ is the cut-off mentioned above. By comparing the expressions 
of the $G_i(s)$ functions in both schemes we can get the equivalent $q_i^{max}$ 
in order to keep the low energy chiral information of the $d_i$ constants, and 
therefore one can use equivalently any of the two procedures.

\begin{figure}[ht]
\centerline{\includegraphics[width=0.6\textwidth]{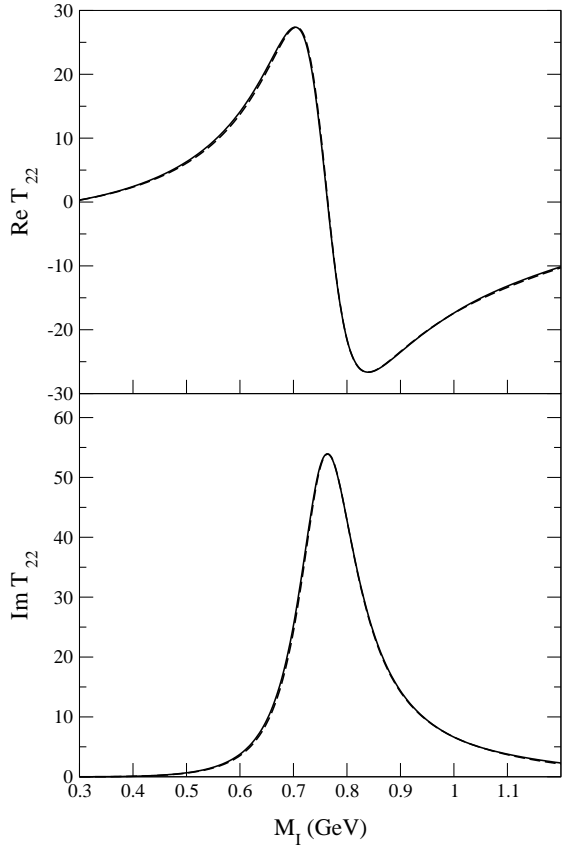}}
\caption{\footnotesize{Real and imaginary parts of the $\pi \pi \to \pi \pi$
scattering amplitude ($T_{22}$ matrix element). The solid line curve is obtained
with the coupled channel formalism including kaons, while the dashed curve is the
result of the decoupled calculation.}}
\label{diagfree}
\end{figure}

In Fig. \ref{diagfree} we show the $T_{22}$ matrix element vs invariant mass. 
To check how important is the inclusion of the $K \bar{K}$ channel we have 
performed the calculation in both coupled and decoupled cases. By 
decoupled case we mean turning off the $K \bar{K} \to \pi \pi$ interaction, what
is easily done setting $K_{12}=0$ in eqs. (\ref{treeK}) and (\ref{T}). The
conclusion is that kaon loops produce minimum changes in the results and
therefore they can be ignored and we can work in a decoupled description in what
concerns the $\pi \pi \to \pi \pi$ scattering amplitude. In the next sections we
shall develop the model for scattering in nuclear matter. Although we formally
work in the general scenario of the coupled channel approach, we perform our 
calculations in the decoupled case.

It is interesting to establish connection with other approaches where tadpole
terms are explicitly kept in the lagrangian. Indeed in \cite{Urban:1998eg} an
explicit term appears in the lagrangian when the minimal coupling is used to
introduce the gauge vector fields. A contact term,
\begin{eqnarray}
\label{Lcontact}
{\cal L}^{(ct)} = g_{\rho}^2 \rho_{\mu} \rho^{\mu} \pi_{+} \pi_{-},
\end{eqnarray}
is obtained where $g_{\rho}$ is connected to our coupling via
\begin{eqnarray}
\label{coupling}
g_{\rho} = - \frac {M_{\rho} G_V}{f^2}.
\end{eqnarray}

\begin{figure}[ht]
\centerline{\includegraphics[width=0.6\textwidth]{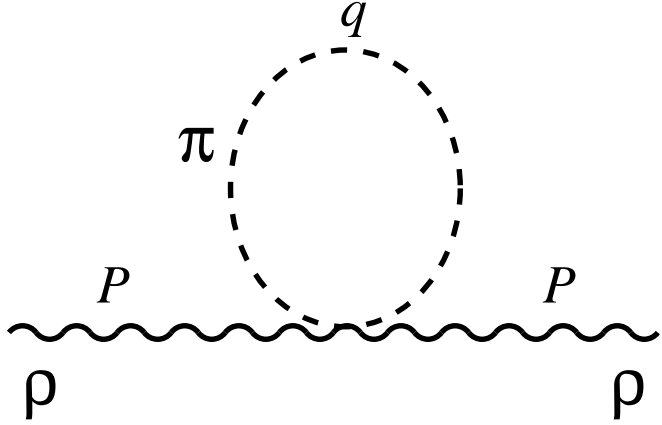}}
\caption{\footnotesize{Tadpole diagram contributing to the $\rho$ meson
selfenergy, coming from the $\rho \rho \pi \pi$ vertex. Wavy lines represent the
$\rho$ meson in the gauge vector field formalism and dashed lines are pions.}}
\label{tadgraph}
\end{figure}

\begin{figure}[ht]
\centerline{\includegraphics[width=0.6\textwidth]{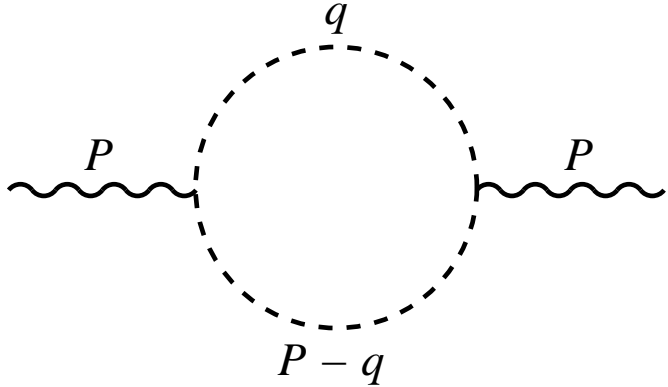}}
\caption{\footnotesize{Two pion diagram contributing to the $\rho$ meson
selfenergy.}}
\label{2pigraph}
\end{figure}

With this lagrangian one obtains a contribution to the $\rho$ selfenergy which
is expressed in terms of the tadpole diagram in Fig. \ref{tadgraph}
and is given by
\begin{eqnarray}
\label{tadpole}
-i t_{tad} = -i \Pi^{\mu \nu} \epsilon_{\mu} \epsilon_{\nu} = -2 g_{\rho}^2 \int
\frac{d^4q}{(2 \pi)^4} g^{\mu \nu} \epsilon_{\mu} \epsilon_{\nu} D_0(q),
\end{eqnarray}
where $D_0 (q)$ is the free pion propagator.
When we work with this formalism we simultaneously release the on shell
condition in the 
$\rho \pi \pi$ vertex in the evaluation of the selfenergy diagram represented in
Fig. \ref{2pigraph} \cite{Oller:1999hw,N/D}.
Since we are interested in the $\rho$ at rest, we need only to evaluate the
spatial components of $\Pi^{\mu \nu}$ and, thus, using the vector form of the
$\rho \pi \pi$ coupling of \cite{Urban:1998eg},
\begin{eqnarray}
\label{rhopipilagrang}
{\cal L}_{\rho \pi \pi} = i g_{\rho} \rho^{\mu} ( \pi^+ \partial_{\mu} \pi^- -
\pi^- \partial_{\mu} \pi^+),
\end{eqnarray}
we obtain
\begin{eqnarray}
\label{PIij}
\Pi^{i j} = i g_{\rho}^2 \int \frac{d^4q}{(2 \pi)^4} 4 q^i q^j D_0 (P-q) D_0(q)
- i g_{\rho}^2 \int \frac{d^4q}{(2 \pi)^4} 2 g^{i j} D_0(q).
\end{eqnarray}
By using dimensional regularization
it is easy to prove that
\begin{eqnarray}
\label{dimtrick}
\int \frac{d^4q}{(2 \pi)^4} \frac{4 q^{\mu} q^{\nu}}{(q^2+s+i \epsilon)^2} =
\int \frac{d^4q}{(2 \pi)^4} \frac{2 g^{\mu \nu}}{q^2+s+i \epsilon}.
\end{eqnarray}
Then eq. (\ref{PIij}) can be written as
\begin{eqnarray}
\label{PIij2}
\Pi^{i j} = i g_{\rho}^2 \int \frac{d^4q}{(2 \pi)^4} 4 q^i q^j \lbrack D_0 (P-q)
D_0(q) - D_0 (q)^2 \rbrack \nonumber \\
= i g_{\rho}^2 \frac{4}{3} \delta^{i j} \int \frac{d^3q}{(2 \pi)^3} \int
\frac{dq^0}{2 \pi} \vec{q}\,^2 \lbrack D_0 (P-q) D_0(q) - D_0 (q)^2 \rbrack
\nonumber \\
= i g_{\rho}^2 \frac{4}{3} \delta^{i j} \int \frac{d^3q}{(2 \pi)^3} \int
\frac{dq^0}{2 \pi} \vec{q}\,^2 \lbrack D_0 (P-q) D_0(q) - \frac{\partial}{\partial
m_{\pi}^2} D_0 (q) \rbrack.
\end{eqnarray}
The $q^0$ integration is now easy to perform and we obtain
\begin{eqnarray}
\label{PIijfinal}
\Pi^{i j} = g_{\rho}^2 \frac{4}{3} \delta^{i j} \int \frac{d^3q}{(2 \pi)^3}
\bigg \lbrack \frac{\vec{q}\,^2}{\omega (q)} \frac{1}{(P^0)^2 - 4 \omega(q)^2 +
i \epsilon} + \frac{\vec{q}\,^2}{\omega (q)} \frac{1}{4 \omega (q)^2} \bigg
\rbrack,
\end{eqnarray}
where $\omega (q)^2 = m_{\pi}^2 + \vec{q}\,^2$.

Now, if we separate in the first term of eq. (\ref{PIijfinal}) the on shell and
off shell contribution of the factor $\vec{q}\,^2$ in the numerator, which comes
from the $\rho \pi \pi$ vertices, we have $\vec{q}\,^2 = \vec{q}\,^2_{on} +
\vec{q}\,^2_{off}$ with $\vec{q}\,^2_{on} = (\frac{P^0}{2})^2 - m_{\pi}^2$.
Since
we have $(P^0)^2 - 4\, \omega (q)^2 = - 4 (\vec{q}\,^2-\vec{q}\,^2_{on})$
then the contribution of the tadpole plus the off shell piece of the two
pion loop diagram is given by
\begin{eqnarray}
\label{offshell}
\tilde{\Pi}^{i j} = - g_{\rho}^2 \frac{m_{\pi}^2}{3} \delta^{i j} \int \frac{d^3q}{(2
\pi)^3} \frac{1}{\omega (q)^3}
\end{eqnarray}
which is a logarithmically divergent piece independent of $P^0$ and hence can be
absorbed in a renormalization of the $\rho$ bare mass.

The former exercise proves that the formalism keeping tadpoles and full off
shell dependence of the $\rho \pi \pi$ vertex is equivalent to the one we have
showed above where only the on shell part of the $\rho \pi \pi$ vertex is kept
and no tadpole is included.

In the medium, however, the pion propagator in the tadpole term will change.
Hence, in order to stick to the gauge invariance of the vector field formalism
we shall keep this tadpole term
\cite{Klingl:1997kf,Herrmann:1993za,Urban:1998eg}.

\section{$(I,J)=(1,1)$ $\pi \pi$ scattering in the nuclear medium}

We now address the calculation in the presence of nuclear matter. As already 
mentioned, the $\rho$ meson couples to intermediate two-meson states. Medium 
corrections will be incorporated in the selfenergies of the mesons involved and
the related vertex corrections as well as from direct coupling to the hadrons in
particle-hole ($p-h$) and delta-hole ($\Delta-h$) excitations. We briefly discuss below the input used for the
pion selfenergy.

\subsection{Pion selfenergy in dense nuclear matter}
\begin{figure}[ht]
\centerline{\includegraphics[width=0.7\textwidth]{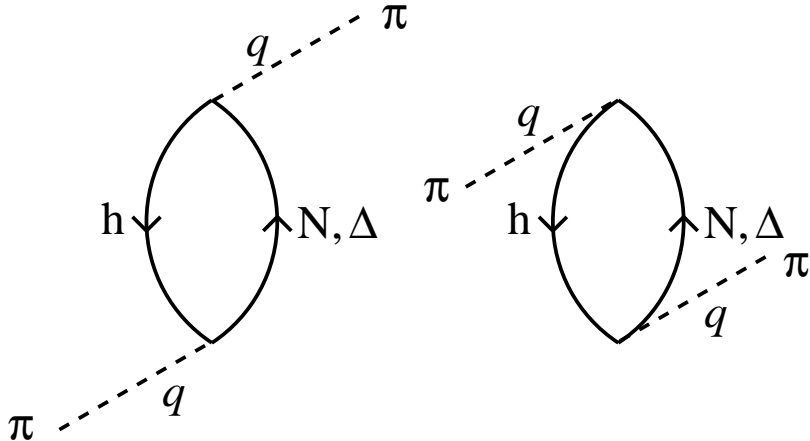}}
\caption{\footnotesize{Direct and crossed $p-h$, $\Delta-h$ excitation terms
giving rise to the pion selfenergy in the nuclear medium, expressed in terms of
the Lindhard function as described in the text.}}
\label{diagph}
\end{figure}

The pion selfenergy in the nuclear medium originates from $p-h$ 
and $\Delta-h$ excitations, corresponding to diagrams in Fig.
\ref{diagph}. We
describe the selfenergy as usual in terms of the ordinary Lindhard function, 
which automatically accounts for
forward and backward propagating bubbles, hence including both diagrams of Fig.
\ref{diagph}. Short range correlations are
also considered with the Landau-Migdal parameter $g'$ set to $0.7$. The final 
expression reads
\begin{eqnarray}
\label{Piself1}
\Pi_{\pi}(q,\rho) = f(\vec{q}\,^2)^2 \vec{q}\,^2 \frac{C \, U(q,\rho)}{1-C \, g'
U(q,\rho)}, 
\end{eqnarray}
where here $q$ is the four-momentum of the pion, $\rho$ stands for the nuclear 
density and $C=(\frac{D+F}{2 f})^2$.
The Lindhard function, 
$U=U_N+U_{\Delta}$, contains both $p-h$ and $\Delta-h$ contributions. 
Formulae for $U_N$ and $U_{\Delta}$
with the normalization required here can be found in the appendix of ref. 
\cite{ReportdeEulogio}. We use a
monopole form factor $f(\vec{q}\,^2)=\frac{\Lambda^2}{\Lambda^2 + \vec{q}\,^2}$
for the $\pi N N$ and $\pi N \Delta$ 
vertices with the cut-off parameter set to $\Lambda=1$ GeV. In the case of the
$\pi N \Delta$ vertex we also include a recoil factor $M_N / \sqrt{s_{\Delta}}$
to account for the fact that the momentum $\vec{q}$ in the $\pi N \Delta$ vertex
should be in $\Delta$ CM frame.
We will check
the sensitivity of our results to variations of both the $\Lambda$ 
and $g'$ parameters.

\subsection{Gauge invariant contact terms}
\begin{figure}[ht]
\centerline{\includegraphics[width=0.7\textwidth]{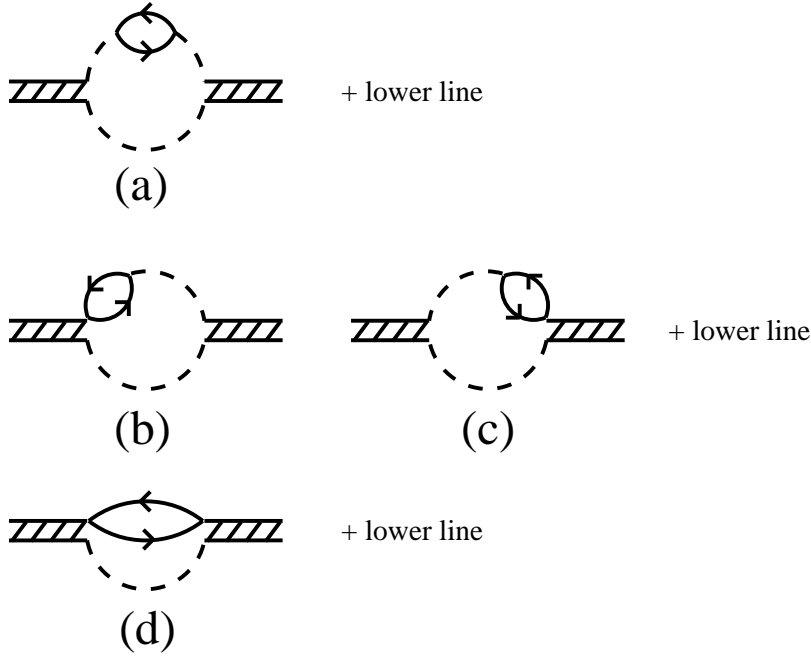}}
\caption{\footnotesize{Medium correction graphs: single solid lines are reserved 
for particle-hole excitations.}}
\label{diagmed}
\end{figure}

\begin{figure}[ht]
\centerline{\includegraphics[width=0.7\textwidth]{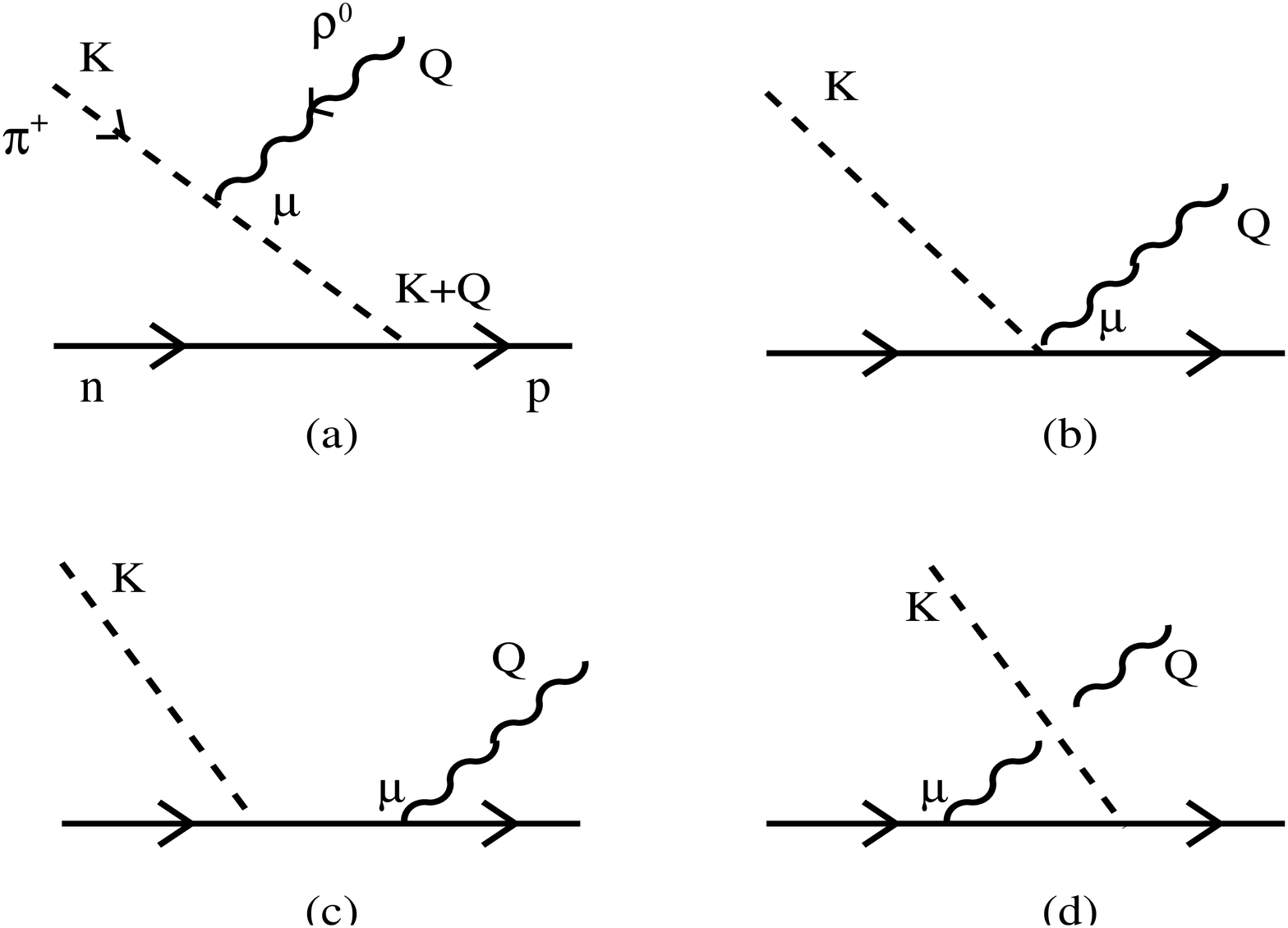}}
\caption{\footnotesize{Gauge invariant set of diagrams used to calculate the 
$\rho$-meson-baryon contact term
as explained in the text. The wavy lines represent here the vector meson, the 
dashed ones pions and the solid lines are reserved for nucleons.}}
\label{diaggauge}
\end{figure}

In our microscopic model of $\pi \pi$ scattering in nuclear matter all graphs in
Fig. \ref{diagmed} are considered. In 
addition to the single $p-h$ excitation (diagram \ref{diagmed}a) we have to 
account for the medium modifications of the $\rho$-meson-meson vertex via the 
$\rho$-meson-baryon contact term, as requested by the gauge invariance of the 
theory \cite{Klingl:1997kf,Chanfray:1993ue,Herrmann:1993za}. Hence we need to
evaluate also the diagrams \ref{diagmed}b to 
\ref{diagmed}d. This contact term can be constructed in a similar way as it 
was done in \cite{Angels} for the $\phi$ decay in the nuclear medium. Let us 
study the set of diagrams depicted in Fig. \ref{diaggauge}, for the case of an
incoming $\pi^+$. The amplitude
corresponding to the graph \ref{diaggauge}a in the gauge 
vector formulation of the $\rho$ is readily
evaluated and its contribution is
\begin{eqnarray}
\label{contact1}
-i t = i \frac{M_{\rho} G_{V}}{f^2} \epsilon_{\mu} (Q+2K)^{\mu} 
\frac{i}{(Q+K)^2-m_{\pi}^2} \sqrt{2}
(\frac{D+F}{2 f}) \vec{\sigma} (\vec{K}+\vec{Q}),
\end{eqnarray}
where $K$ and $Q$ are four-momenta of the pion and the $\rho$ meson 
respectively, $\vec{\sigma}$ is the spin-$\frac{1}{2}$ operator
and $\epsilon_{\mu}$ is the polarization vector of the $\rho$ meson. The contact
term, represented by Fig. \ref{diaggauge}b, must have the structure
\begin{eqnarray}
\label{contact2}
-i t_{cont} = \epsilon_{\mu} B^{\mu},
\end{eqnarray}
and gauge invariance requires that if we replace $\epsilon_{\mu}$ by $Q_{\mu}$ 
then the sum of all the terms in Fig. \ref{diaggauge} vanishes. To find out the
strength of the coupling we shall take the limit $\vec{K} \to \vec{0}$ 
because in that case diagrams 
\ref{diaggauge}c and \ref{diaggauge}d give a null contribution and we have to 
consider only the sum of the first two so far commented. Then
we find the following system of equations:
\begin{eqnarray}
\label{contact3}
B^{0} = 0 \nonumber \\
-\frac{M_{\rho} G_{V}}{f^2} \sqrt{2} (\frac{D+F}{2 f}) \vec{\sigma} \vec{Q} - 
\vec{B} \vec{Q} = 0.
\end{eqnarray}
After solving for $B^{\mu}$, the amplitude corresponding to diagram 
\ref{diaggauge}b is
\begin{eqnarray}
\label{contact4}
-i t_{cont} = \frac{M_{\rho} G_{V}}{f^2} \sqrt{2} (\frac{D+F}{2 f}) \vec{\sigma} 
\vec{\epsilon},
\end{eqnarray}
and with this information the whole set of diagrams in Fig. \ref{diagmed} can be 
evaluated. Note that because of the $\rho$ coupling to the pions, eq.
(\ref{contact4}) has a minus sign for an incoming $\pi^-$.

\subsection{Technical implementation of vertex corrections} 
In the previous section we have commented that gauge invariance requires
the presence of the $\rho$-meson-baryon contact
term, which results in the need of including diagrams b, c and d in Fig.
\ref{diagmed}. Now we pay attention on how these vertex
corrections can be incorporated in the unitary treatment of the $T$ matrix 
described in section 2. To this end let us
\begin{figure}[ht]
\centerline{\includegraphics[width=0.7\textwidth]{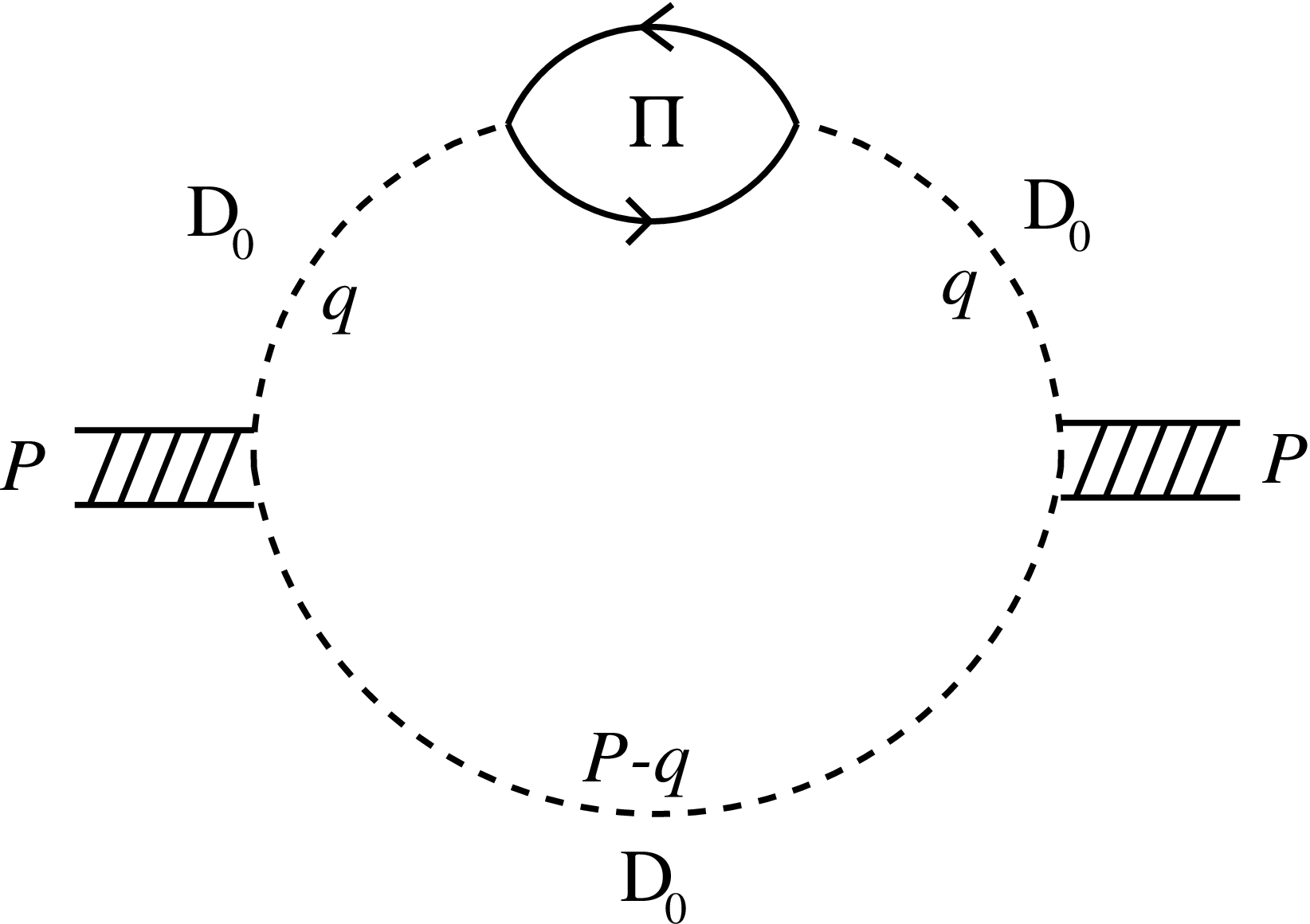}}
\caption{\footnotesize{Momentum labels for the in-medium correction diagrams. 
$D_0$ indicates a pion propagator and
$\Pi$ refers to the pion selfenergy due to $p-h$ excitation.}}
\label{diagdetail}
\end{figure}
consider in detail the set of diagrams \ref{diagmed}a, b and c \footnote{We
restrict here the discussion to the case of pions. The argument for kaons would
be analogous.}. The momentum
labelling is shown in Fig. \ref{diagdetail} for the first diagram
(the pion in the upper line will have four-momentum $q$ and the
one in the lower line ($P-q$)). The amplitudes originating from
these graphs in the gauge vector formulation of the $\rho$ are evaluated using
Feynman rules and the results are
\begin{eqnarray}
\label{vcamplitudes}
A_{(a)} &=& \int {\frac{d^{4}q}{(2\pi)^{4}} V_{\rho \pi \pi}(q)^{2} iD_0(q) iD_0(P-q) 
\bigg \lbrack -i (\frac{D+F}{2 f})^2 \vec{q}\,^2 U(q,\rho) \bigg \rbrack iD_0(q)} \nonumber \\
A_{(b)} &=& \int {\frac{d^{4}q}{(2\pi)^{4}} V_{\rho \pi \pi}(q)^{2} 
\frac{1}{2} \bigg \lbrack(\frac{D+F}{2 f})^2 U(q,\rho) \bigg \rbrack iD_0(q) iD_0(P-q)}
\nonumber \\
A_{(c)} &=& A_{(b)}
\end{eqnarray}
where $V_{\rho \pi \pi}$ stands for the $\rho \pi \pi$ vertex,
\begin{eqnarray}
\label{rhopipivertex}
V_{\rho \pi \pi} = -i \frac{M_{\rho} G_V}{f^2} 2 \vec{\epsilon} \vec{q}
\end{eqnarray}
and $D_0(q)$ is the bare propagator of a pion with four-momentum $q$. The
subindices (a), (b) and (c) refer to each
diagram in Fig. \ref{diagmed}. In $A_{(a)}$ we have explicitly substituted the
expression of the pion selfenergy due to a $p-h$ or $\Delta -h$ excitation, 
which would read
\begin{eqnarray}
\label{self_oneph}
-i \Pi(q,\rho) = -i (\frac{D+F}{2 f})^2 \vec{q}\,^2 U(q,\rho).
\end{eqnarray}

It is obvious that the sum of all three diagrams,
\begin{eqnarray}
\label{vcsum}
\lefteqn{A_{(a)} + A_{(b)} + A_{(c)} = {} } \nonumber \\
& & {}\int {\frac{d^{4}q}{(2\pi)^{4}} V_{\rho \pi \pi}(q)^{2} 
iD_0(q) iD_0(P-q) \bigg \lbrack(\frac{D+F}{2 f})^2 U(q,\rho)
\bigg \rbrack (\vec{q}\,^2 D_0(q) + 1)},
\end{eqnarray}
can be cast in the form of $A_{(a)}$ if we substitute the $\vec{q}\,^2$
appearing in eq. (\ref{self_oneph}) by
\begin{eqnarray}
\label{vc1}
\vec{q}\,^2 \to \vec{q}\,^2 + D_0^{-1}(q).
\end{eqnarray}
We still have to consider the last diagram in Fig. \ref{diagmed},
\ref{diagmed}d. It is easy to find, in the same way as before, that the
amplitude can be cast in the structure of $A_{(a)}$ making the substitution
\begin{eqnarray}
\label{vc2}
\vec{q}\,^2 \to  \frac{3}{4} \frac{D_0^{-2}(q)}{\vec{q}\,^2}
\end{eqnarray}
in eq. (\ref{self_oneph}). Hence, we find that vertex corrections can be
automatically introduced in the gauge vector formulation of the problem by
performing the following substitution
\begin{eqnarray}
\label{vcfinal}
\vec{q}\,^2 \to H_{vc}(q) \equiv \vec{q}\,^2 + D_0^{-1}(q) + \frac{3}{4}
\frac{D_0^{-2}(q)}{\vec{q}\,^2}
\end{eqnarray}
in the pion selfenergy described in eq. (\ref{Piself1}),
hence working with a kind of an 'extended' pion selfenergy which takes care of
the medium modifications of the $\rho \pi \pi$ vertex, namely
\begin{eqnarray}
\label{selfextended}
-i \Pi^{ext}(q,\rho) = -i f(\vec{q}\,^2)^2 H_{vc}(q) \frac{(\frac{D+F}{2 f})^2
U(q,\rho)}{1 - (\frac{D+F}{2 f})^2 g' U(q,\rho)}.
\end{eqnarray}

The calculation of the $\pi\pi$ scattering matrix in the medium follows the
calculation of the $G(s)$ function for pions. As it was mentioned in section 2 we
keep full off shell dependence of the $\rho\pi\pi$ vertex. The medium effects
are introduced by a subtraction of the $\rho$ meson selfenergy due to pion
loops in free space from the same quantity calculated in nuclear matter. The
following substitution is performed in eq. (\ref{g(s)}):
\begin{eqnarray}
\label{gtogmedio}
G(s) \to G(s) + \frac{1}{\vec{q}\,^2_{on}} \lbrack I_{med}(s) - I_{free}(s)
\rbrack,
\end{eqnarray}
where
\begin{eqnarray}
\label{Ifunctions}
I_{med}(s) &=& i \int \frac{d^{4}q}{(2\pi)^{4}} \vec{q}\,^2 D(q) D(P-q)
\nonumber\\
I_{free}(s) &=& i \int \frac{d^{4}q}{(2\pi)^{4}} \vec{q}\,^2 D_0(q) D_0(P-q)
\end{eqnarray}
and $D(q)$ is the in-medium pion propagator,
\begin{eqnarray}
\label{piprop}
D(q) = \frac{1}{q^2 - m_{\pi}^2 - \Pi_{\pi}(q,\rho)}.
\end{eqnarray}
To include the vertex corrections discussed above, the pion selfenergy
$\Pi_{\pi}(q,\rho)$ in the latter equation must be replaced by the one in eq.
(\ref{selfextended}).

Each of the integrals in eq. (\ref{gtogmedio}) are quadratically divergent. The
subtraction $I_{med}-I_{free}$ cancels these quadratic divergences and we are
left with a logarithmic divergence which in principle has to be regularized with
a cut off in the pion loop momentum. However, notice that $D(q) D(q) - D_0 (q)
D_0 (q)$ is proportional to $\Pi^{ext}(q,\rho)$. Therefore,
since all the pieces that enter
$\Pi^{ext}(q,\rho)$ are multiplied by a squared form factor $f(\vec{q}\,^2)^2$
the subtraction is convergent.

\subsection{Rho meson propagator from the T matrix}
For the sake of simplicity, and having in mind the results of section 2, we
turn off the interaction between $\pi \pi$ and $K \bar{K}$ systems and
work in a decoupled channel approach ($K_{12}=0$ in eqs. (\ref{treeK}) and
(\ref{T})). Then the $\pi \pi \to \pi \pi$ $T$ matrix element takes the simple 
form
\begin{eqnarray}
\label{T22dec}
T_{22} = \frac{1}{1+K_{22} G} K_{22}
\end{eqnarray}
Let us first assume that $\frac{2 G_V^2}{f^2}=1$, as it is indeed the case if VMD holds exactly. Then the $K_{22}$
matrix element reads
\begin{eqnarray}
\label{K22VMD}
K_{22} = - \frac{2}{3} \frac{p_{\pi}^2 M_{\rho}^2}{f^2} \frac{1}{s-M_{\rho}^2},
\end{eqnarray}
where one can identify the p-wave structure of the $\pi \pi \to \pi \pi$
scattering amplitude mediated by the exchange of a $\rho$ meson.
Hence the factor $(1+K_{22} G)^{-1}$ in eq. (\ref{T22dec}) is accounting for a selfenergy of the $\rho$ meson 
arising from the
coupling with $\pi \pi$ pairs (to switch on the presence of the nuclear medium
one has just to perform the substitution of eq. (\ref{gtogmedio}) as described in section 3 with the
modified pion selfenergy of 
eq. (\ref{selfextended})). We can therefore write eq. (\ref{T22dec}) as
\begin{eqnarray}
\label{T22decexp}
T_{22} &=& - \frac{2}{3} \frac{p_{\pi}^2 M_{\rho}^2}{f^2} \frac{1}{s-M_{\rho}^2-\Pi_{\rho}^{\pi-loop}} \nonumber \\
\Pi_{\rho}^{\pi-loop} &=& - K_{22} G (s-M_{\rho}^2) \,=\, \frac{2}{3}
\frac{p_{\pi}^2 M_{\rho}^2}{f^2} G,
\end{eqnarray}
where $\Pi_{\rho}^{\pi-loop}$ is the selfenergy commented above.

Now we consider the case $\frac{2 G_V^2}{f^2} \ne 1$. Let $a \equiv \frac{2 G_V^2}{f^2}$ and proceed as
above. The $K_{22}$ matrix element can be easily cast in the form
\ba
\label{K22noVMD}
K_{22} = - \frac{2}{3} \frac{p_{\pi}^2 M_{\rho}^2}{f^2} B(s)
\frac{1}{s-M_{\rho}^2} 
\ea
where $B(s) \equiv 1+ (a-1) \frac{s}{M_{\rho}^2}$. Eq. (\ref{K22noVMD}) reduces to eq. (\ref{K22VMD}) if $a=1$, because
in that case $B=1$ independently of $s$. 
For the values of $G_V$ and $f$ of \cite{Oller:2001ug} we find that $a$ takes
the value $\approx 0.7$.
For instance, at $s=M_{\rho}^2$ we have $B(M_{\rho}^2) \approx 0.7$, therefore a 30 \% of
deviation from VMD is introduced. The $T$ matrix element for $\pi \pi \to \pi \pi$ scattering now reads
\begin{eqnarray}
\label{T22decexpB}
T_{22} &=& - \frac{2}{3} \frac{p_{\pi}^2 M_{\rho}^2}{f^2} B(s)
\frac{1}{s-M_{\rho}^2-\Pi_{\rho}^{\pi-loop}} = - \tilde{g}^2 \frac{4}{3}
p_{\pi}^2 D_{\rho} (s)
\nonumber \\ 
\Pi_{\rho}^{\pi-loop} &=& \frac{2}{3} \frac{p_{\pi}^2 M_{\rho}^2}{f^2} B(s) G
\,=\, \tilde{g}^2 \frac{4}{3} p_{\pi}^2 G,
\end{eqnarray}
where $\tilde{g}^2 = \frac{1}{2} \frac{M_{\rho}^2}{f^2} B(s)$ defines a coupling
which will replace $g_{\rho}^2$ in our formulae, and $D_{\rho}$ is the $\rho$
meson propagator.

In the following sections we shall discuss about the tadpole term in the nuclear
medium as well as new contributions due to the direct coupling of the $\rho$
meson to the baryonic lines in particle and delta-hole excitations. These new
contributions will manifest as additional pieces to the whole $\rho$ meson
selfenergy $\Pi$. The way of introducing these new pieces in the $T$ matrix and
the $\rho$ propagator follows the same scheme as it has been done with the
contribution from two pion loop selfenergy plus its vertex corrections,
$\Pi_{\rho}^{\pi -loop}$.

\subsection{Tadpole term in nuclear matter}
We have seen that keeping the full off shell dependence of the $\rho \pi \pi$
vertex and explicitly considering the free tadpole term in the calculation of
the $\rho$ meson selfenergy is equivalent to calculating with the on shell part
of the $\rho \pi \pi$ vertex and describing the pion loop contributions with the
$G(s)$ functions defined in section 2. This freedom allows us to include the
tadpole diagram in nuclear matter by performing a subtraction between the
selfenergies corresponding to the tadpole term in the medium and in free space.
In the gauge vector field approach one has
\begin{eqnarray}
\label{tad1}
-i \Pi_{\mu\nu}^{tad} = -2 g_{\rho}^2 g_{\mu\nu} \int \frac{d^{4}q}{(2\pi)^{4}}
D_0 (q).
\end{eqnarray}
The $\rho$ selfenergy, $\Pi$, which appears in the $\rho$ propagator in the
former section can be related to $\Pi_{\mu\nu}$ by
\begin{eqnarray}
\label{PimunuvsPi}
\Pi_{\mu\nu} = \bigg \lbrace - g_{\mu\nu} + \frac{P^{\mu} P^{\nu}}{M_{\rho}^2}
\bigg \rbrace \Pi,
\end{eqnarray}
and restricting ourselves to the space components of $\mu$, $\nu$ we have for
the tadpole piece
\begin{eqnarray}
\label{tad1b}
\Pi^{tad} = 2 i g_{\rho}^2 \int \frac{d^{4}q}{(2\pi)^{4}} D_0 (q).
\end{eqnarray}
We define the subtracted in medium tadpole contribution to the $\rho$ meson
selfenergy as
\begin{eqnarray}
\label{tadsub1}
\tilde{\Pi}^{tad} (\rho) = 2 i g_{\rho}^2 \int \frac{d^{4}q}{(2\pi)^{4}} \lbrack
D (q) - D_0 (q) \rbrack.
\end{eqnarray}

Given the analytical structure of $D(q)$ and $D_0(q)$ by performing a Wick
rotation one can see that $\tilde{\Pi}^{tad}$ is real, and eq. (\ref{tadsub1})
can be rewritten as
\begin{eqnarray}
\label{tadsub1b}
\tilde{\Pi}^{tad} (\rho) = - 2 g_{\rho}^2 \int \frac{d^{4}q}{(2\pi)^{4}} \lbrack
Im D (q) - Im D_0 (q) \rbrack.
\end{eqnarray}
Taking into account that both $D_0 (q)$ and $D (q)$ are even functions of $q^0$,
the integration in $q^0$ can be replaced by twice the integration from $0$ to
$\infty$. This integration can be performed analytically for the second term of
eq. (\ref{tadsub1b}) and we obtain
\begin{eqnarray}
\label{tadsub1c}
\tilde{\Pi}^{tad} (\rho) = - 4 g_{\rho}^2 \int \frac{d^{3}q}{(2\pi)^{3}} \bigg
\lbrack \int_0^{\infty} \frac{dq^0}{2\pi} Im D (q) + \frac{1}{4 \omega(q)} \bigg \rbrack.
\end{eqnarray}
Eq. (\ref{tadsub1c}) is free from quadratic divergences. Actually the form
factors present in the pion selfenergy make it convergent and a cut off in the
pion loop momentum is not required.

\subsection{Other medium corrections}

As it has been discussed in subsections 3.2 and 3.3, gauge invariance provides us
with $\rho \pi N N$, $\rho \pi N \Delta$ contact vertices which give rise to the
diagrams depicted in Fig. \ref{diagmed}. In addition to those there are
other medium corrections that arise within the gauge vector field
formalism and generate interactions via the minimal coupling scheme (the former
contact vertices can also be obtained in this way). Let us summarize the
interaction lagrangians of ref. \cite{Urban:1998eg} involving $\rho \pi \pi$,
$\rho \rho \pi \pi$, $\pi N N$, $\pi N \Delta$, $\rho N N$, $\rho \pi N N$,
$\rho \Delta \Delta$ and $\rho \pi N \Delta$ vertices:

\begin{eqnarray}
\label{urbanlagrangians}
{\cal L}_{\pi\rho} &=& {1\over 2} ig_{\rho} \rho_{\mu} (T_3 \vec{\phi}
\partial^{\mu}\vec{\phi} + \partial^{\mu}\vec{\phi} T_3\vec{\phi})
- {1\over 2} g_{\rho}^2 \rho_{\mu} \rho^{\mu} T_3 \vec{\phi} T_3 \vec{\phi} \nonumber\\
{\cal L}_{\pi N N} &=& {f_N \over {m_{\pi}}}\, \bar{\psi}\gamma^5\gamma^{\mu}
                   \vec{\tau}\psi \partial_{\mu}\vec{\phi} \nonumber\\
{\cal L}_{\pi N\Delta} &=& -{f_{\Delta}\over {m_{\pi}}}
  \bar{\psi}\vec{T}^{\dag}\psi_{\mu}\partial^{\mu}\vec{\phi}
 \quad +\quad\mbox{h.c.} \nonumber\\
{\cal L}_{\rho N N}&=&-{g_{\rho}\over 2}\bar{\psi}\rhoslash \tau_3 \psi \nonumber\\
{\cal L}_{\rho\pi N N}&=&ig_{\rho}{f_N \over {m_{\pi}}} \bar{\psi}\gamma^5\rhoslash
   \vec{\tau}\psi T_3\vec{\phi} \nonumber\\
{\cal L}_{\rho\Delta\Delta} &=&
  g_{\rho}\bar{\psi}_{\mu}\rhoslash T_3^{({3\over 2})}\psi^{\mu}
  -{g_{\rho}\over 3}\bar{\psi}_{\mu}(\gamma^{\mu}\rho_{\nu}+\gamma_{\nu}\rho^{\mu})
  T_3^{({3\over 2})}\psi^{\nu}
  +{g_{\rho}\over 3}\bar{\psi}_{\mu}\gamma^{\mu}\rhoslash T_3^{({3\over 2})}
  \gamma_{\nu}\psi^{\nu} \nonumber\\
{\cal L}_{\rho\pi N\Delta} &=&
  -ig_{\rho}{f_{\Delta}\over{m_{\pi}}}\bar{\psi}\vec{T}^{\dag}\psi_{\mu}\rho^{\mu}
  T_3\vec{\phi}\quad +\quad\mbox{h.c.}
\end{eqnarray}

From the list above, lagrangians generating direct couplings of a $\rho$ meson
to either a nucleon or a delta were not considered in subsection 3.2. As we will
discuss below, these terms give rise to a set of diagrams correcting both the
$\rho$ meson selfenergy from two pion loop and tadpole terms.

\begin{figure}[ht]
\centerline{\includegraphics[width=0.6\textwidth]{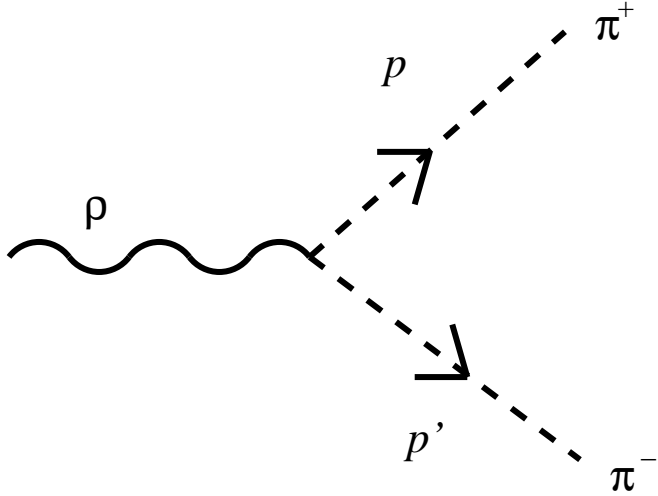}}
\caption{\footnotesize{$\rho \pi \pi$ vertex. Lines as in Fig. \ref{diaggauge}}}
\label{rhopipivfig}
\end{figure}

Before proceeding with the discussion of the new contributions, it is
interesting to comment on the relative sign of the $\rho \pi \pi$ coupling
constant in the language of chiral lagrangians and in the gauge vector field
formalism. From eq. (\ref{urbanlagrangians}), the $\rho \pi \pi$ vertex takes
the form
\begin{eqnarray}
\label{rhopipiurban}
-i t_{\rho\pi\pi} = - i g_{\rho} \epsilon_{\mu} (p - p')^{\mu},
\end{eqnarray}
with the momentum labels corresponding to Fig. \ref{rhopipivfig}.
However, in our model we would write
\begin{eqnarray}
\label{rhopipinuestro}
-i t_{\rho\pi\pi} = i \frac{M_{\rho} G_V}{f^2} \epsilon_{\mu} (p - p')^{\mu},
\end{eqnarray}
and this establishes the correspondence
\begin{eqnarray}
\label{couplings}
g_{\rho} = - \frac{M_{\rho} G_V}{f^2}
\end{eqnarray}
to connect both schemes. As an example, following the lagrangians listed above,
eq. (\ref{contact4}) reads
\begin{eqnarray}
\label{contact4_urban}
-i t_{cont} = -g_{\rho}  {f_N \over {m_{\pi}}} \sqrt{2} \vec{\sigma} \vec{\epsilon},
\end{eqnarray}
which reconfirms the relationship of eq. (\ref{couplings}).

\subsubsection{Extra two-pion loop correction to the $\rho$ selfenergy}
\begin{figure}[ht]
\centerline{\includegraphics[width=0.6\textwidth]{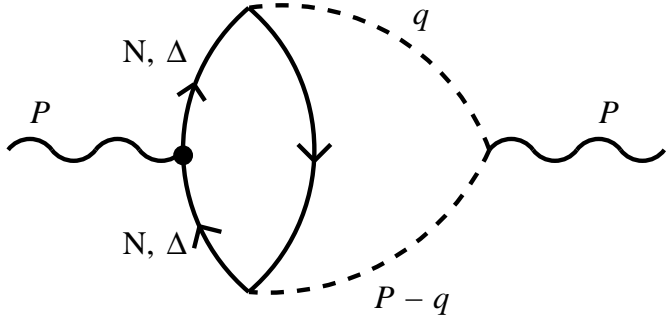}}
\caption{\footnotesize{Vertex correction involving $\rho NN$, $\rho \Delta
\Delta$ vertices.}}
\label{rhoNNcorrection}
\end{figure}

The first additional medium correction that we consider here is the one shown in
Fig. \ref{rhoNNcorrection}, with the $\rho$ meson attached to a particle line.
As we shall show below, this contribution can also be cast in the form of a
vertex correction to be introduced in the 'extended' pion selfenergy of eq.
(\ref{selfextended}). Following the Feynman rules for the vertices involved
(after a non-relativistic reduction) the $\rho$
meson selfenergy contribution from this diagram ($N-h$ case, $P^0=\sqrt{s}$)
reads
\begin{eqnarray}
\label{PirhoNNcorr}
-i \Pi_{1,\rho N N} ^{ij} &=& \delta ^{ij} g_{\rho}^2 \frac{4}{3} \frac{1}{M_N
\sqrt{s}} \bigg ({f_N
\over {m_{\pi}}} \bigg )^2 \int \frac{d^{4}q}{(2\pi)^{4}} (\vec{q}\,^2)^2 D_0
(q) D_0(P-q){} \nonumber\\
& & {}\frac{1}{2} \lbrack \bar{U}_N (q^0-\sqrt{s},\vec{q}) - \bar{U}_N
(q^0,\vec{q}) + \bar{U}_N (-q^0,\vec{q}) - \bar{U}_N (-q^0+\sqrt{s},\vec{q})
\rbrack, 
\end{eqnarray}
where the $\bar{U}_N$ functions account for the direct contribution of a
nucleon-hole excitation (Fig. \ref{diagph}, left) and sum over spin but not over
isospin. They satisfy the relation
\begin{eqnarray}
\label{UbvsU}
2 \, \lbrack \bar{U}_N (q^0,\vec{q}) + \bar{U}_N (-q^0,\vec{q}) \rbrack = 
U_N (q^0,\vec{q}).
\end{eqnarray}

In the case of a $\rho$ meson coupling to a $\Delta$ line one obtains a similar
expression, with different spin/isospin factors and a scaling factor to the
$\pi\pi N$ coupling, if a constant $\Delta$ width is considered. The
reason is that in the nucleon case one is allowed to express the product of two
particle propagators as the difference of such propagators times
$s^{-\frac{1}{2}}$. Then the resulting expression of $\Pi ^{ij}$ can be
written in terms of the $\bar{U}_N$ functions. However, keeping the $\Delta$
width is important and, hence, we shall keep its energy
dependence and use another approximation to the integral of the
two $\Delta$ and one hole propagators appearing in the expression of $\Pi^{ij}$.
The calculation is the done by making a Fermi average over the two $\Delta$
propagators.

Given the structure of this correction it is convenient to incorporate it by
means of corrections to the pion selfenergy in the two pion propagators of eq.
(\ref{Ifunctions}). This is accomplished by means of the substitution
\begin{eqnarray}
\label{lastVC}
\vec{q}\,^2 U(q^0,\vec{q}) \to \vec{q}\,^2 U(q^0,\vec{q}) + 
(\alpha_N + \alpha_{\Delta}) D_0^{-1}(q^0,\vec{q}),
\end{eqnarray}
with
\begin{eqnarray}
\label{alphaybeta}
\alpha_N &=& \frac{\vec{q}\,^2}{2 M_N \sqrt{s}} \lbrack \bar{U}_N
(q^0-\sqrt{s},\vec{q}) - \bar{U}_N (q^0,\vec{q}) + 
\bar{U}_N (-q^0,\vec{q}) - \bar{U}_N (-q^0+\sqrt{s},\vec{q}) \rbrack \nonumber\\
\alpha_{\Delta} &=& \frac{\vec{q}\,^2}{2 M_{\Delta}} \frac{10}{9}
\bigg ( \frac{f_{\Delta}}{f_N} \bigg )^2 \frac{\rho}{2} \bigg \lbrace
G_{\Delta}(q^0-\sqrt{s},\vec{q}) G_{\Delta}(q^0,\vec{q}) +{} \nonumber\\
& & {}G_{\Delta}(-q^0,\vec{q}) G_{\Delta}(-q^0+\sqrt{s},\vec{q}) \bigg \rbrace
\end{eqnarray}
and
\begin{eqnarray}
\label{Gdelta}
G_{\Delta}(q^0,\vec{q}) = \frac{1}{\sqrt{s_{\Delta}}-M_{\Delta}+\frac{i}{2}
\Gamma_{\Delta}(s_{\Delta})} \nonumber \\
s_{\Delta} = M_N^2 + (q^0)^2 - \vec{q}\,^2 + 2 q^0 \sqrt{M_N^2 + \frac{3}{5}
k_F^2}.
\end{eqnarray}
Expressions for the energy dependent $\Delta$ width can be found in
\cite{ReportdeEulogio}.The substitution of eq. (\ref{lastVC}) already accounts for the two possible
time orderings in Fig. \ref{rhoNNcorrection}: incoming $\rho$ meson coupled to
the baryon loop, outgoing $\rho$ meson closing the two pion loop, and vice versa.
Notice that in eq. (\ref{Gdelta}) we have not used the common non-relativistic
approximation of $G_{\Delta}^{-1}$ since it is not appropriate for this case
because high momenta contribute to the loop integrals. The same
argument is of application for the corrections discussed in the next section.

\subsubsection{Tadpole like corrections}
\begin{figure}[ht]
\centerline{\includegraphics[width=0.6\textwidth]{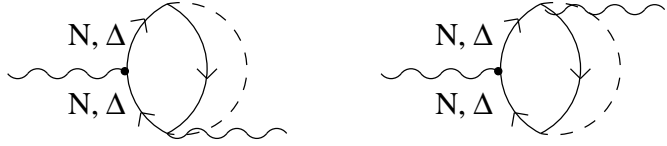}}
\caption{\footnotesize{Additional medium corrections. The pion line
represent either a $\pi^+$ or a $\pi^-$. The partners with the outgoing $\rho$
attached to the nucleon (or $\Delta$) line are not displayed.}}
\label{tadcorrection}
\end{figure}

We study now a set of diagrams containing $\rho N N$, $\rho\Delta\Delta$ vertices
which are generated from those in the former section by contracting one pion
line, in analogy to the way the vertex corrections in Figs. \ref{diagmed}b,
\ref{diagmed}c where generated from Fig. \ref{diagmed}a. These terms are related
to the diagrams labelled as $\rho \rho \pi \pi$ vertex correction in
\cite{Urban:1998eg}. The diagrams are shown in Fig. \ref{tadcorrection},
although the set includes also the partners with an outgoing $\rho$ meson
attached to a nucleon (or $\Delta$) line.

The selfenergy corresponding to the diagrams in Fig. \ref{tadcorrection} and
their partners for the nucleon case reads
\begin{eqnarray}
\label{tadcorr1}
-i \Pi_{2,\rho N N}^{ij} = \frac{4}{3} \frac{g_{\rho}^2}{M \sqrt{s}} \bigg ({f_N
\over {m_{\pi}}} \bigg )^2 \delta^{ij} \int \frac{d^{4}q}{(2\pi)^{4}}
\vec{q}\,^2 D_0 (q) \lbrack \bar{U}_N (q^0-\sqrt{s},\vec{q}) - \bar{U}_N
(q^0+\sqrt{s},\vec{q}) \rbrack,
\end{eqnarray}
where the factor four indicates that the contributions of protons or neutrons
over the Fermi sea are identical, and the same happens for the diagrams with the
incoming and outgoing $\rho$ mesons interchanged.

For the case of the $\Delta$ isobar, the number of diagrams involved are twice
the number for the nucleon case, since now the $\Delta$ can be excited to its
four charge states. As it happened in the previous subsection the result cannot
be expressed in terms of modified Lindhard functions. Instead, we use the same
approximation as before for the integration of the fermionic loop and we get
\begin{eqnarray}
\label{tadcorr2}
-i \Pi_{2,\rho\Delta\Delta}^{ij} &=& \frac{4}{3} \frac{g_{\rho}^2}{M_{\Delta}}
\frac{10}{9} \bigg ({f_N \over {m_{\pi}}} \bigg )^2 \bigg ({f_{\Delta} \over
f_N} \bigg )^2 \delta^{ij} \int \frac{d^{4}q}{(2\pi)^{4}} 
\vec{q}\,^2 D_0 (q){} \nonumber\\
& & {}\frac{\rho}{2} \bigg \lbrace G_{\Delta}(q^0+\sqrt{s},\vec{q})
G_{\Delta}(q^0,\vec{q}) + G_{\Delta}(q^0,\vec{q})
G_{\Delta}(q^0-\sqrt{s},\vec{q}) \bigg \rbrace.
\end{eqnarray}

\subsubsection{Hartree like nucleon tadpole term}
In ref. \cite{Urban:1998eg} we find no other medium corrections to the
$\rho\pi\pi$ and $\rho\rho\pi\pi$ vertex functions, since the $\rho$ coupling to
baryons is treated in a non-relativistic approach up to order $(1/M_N)^0$. At
next order an additional four point $\rho\rho N N$ vertex enters the
calculation. This can be seen in ref. \cite{Herrmann:1993za} where this vertex
is generated by minimal substitution in a non-relativistic treatment of the
$\rho N N$ lagrangian.

\begin{figure}[ht]
\centerline{\includegraphics[width=0.9\textwidth]{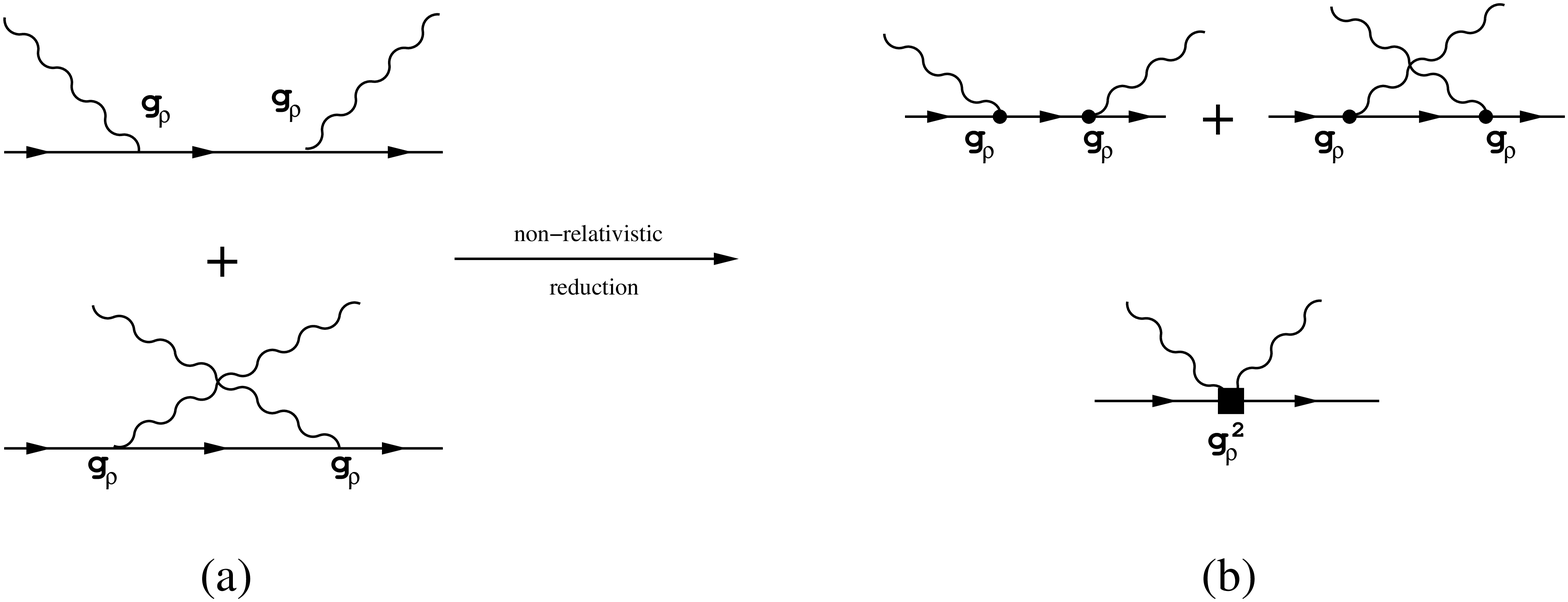}}
\caption{\footnotesize{Diagrammatic representation of the non-relativistic
reduction of the $\rho N \to \rho N$ amplitude.}}
\label{rhoNrhoNamplitude}
\end{figure}

The same result can be obtained by calculating the $\rho N \to \rho N$ amplitude
in the relativistic theory and then performing the non-relativistic reduction up
to order $M_N^{-1}$. This is schematically shown in Fig.
\ref{rhoNrhoNamplitude}. In this non-relativistic reduction we obtain the
nucleon direct and crossed pole terms of Fig. \ref{rhoNrhoNamplitude}b with
non-relativistic propagators and vertices, plus a contact term of order
$M_N^{-1}$.

\begin{figure}[ht]
\centerline{\includegraphics[width=0.7\textwidth]{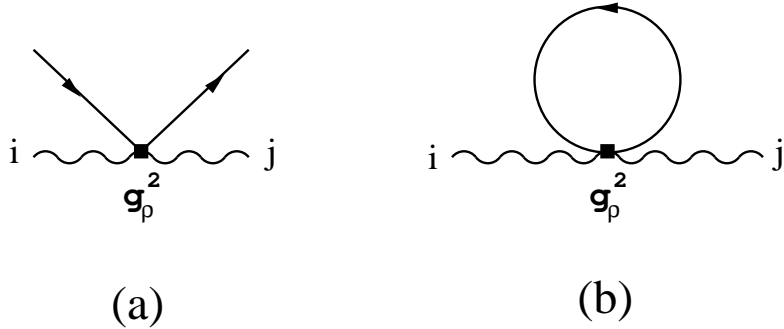}}
\caption{\footnotesize{(a) $\rho \rho N N$ vertex. (b) Nucleon tadpole term.}}
\label{rhoNrhoNtadpole}
\end{figure}

If the nucleon lines of this last diagram are closed one obtains a nucleon
tadpole diagram (Fig. \ref{rhoNrhoNtadpole}b) which contributes as an additional
linear-in-density piece of the $\rho$ meson selfenergy \cite{Herrmann:1993za}.
The Feynman rule for the $\rho \rho N N$ vertex (Fig. \ref{rhoNrhoNtadpole}a)
can be derived from the $\rho N \to \rho N$ amplitude and it reads
\begin{equation}
\label{rhoNrhoNvertex}
-i t_{\rho \rho N N}^{ij} = -i g_{\rho}^2 \frac{1}{4} \frac{1}{M_N} \delta
^{ij},
\end{equation}
and the nucleon tadpole term is easily evaluated,
\begin{eqnarray}
\label{rhoNrhoNselfenergy}
-i \Pi_{\rho \rho N N}^{ij} = - \delta ^{ij} g_{\rho}^2 \frac{1}{M_N} \int
\frac{d^4 p}{(2\pi)^4} \frac{n(\vec{p})}{p^0 - \mathcal{E}_N (\vec{p}) - i
\epsilon}
\nonumber \\
= -i \delta ^{ij} g_{\rho}^2 \frac{1}{M_N} \int \frac{d^3 p}{(2\pi)^3}
n(\vec{p}) = -i \delta ^{ij} g_{\rho}^2 \frac{1}{4} \frac{1}{M_N} \rho .
\end{eqnarray}
The result is an energy independent function proportional to the nuclear matter
density. Its inclusion as a selfenergy term in the $\rho$ meson propagator
follows the same procedure as already discussed for other contributions.

Finally, some other medium corrections involving $\rho B B$, $\rho \pi B B'$ and
$\rho \rho B B$ vertices can be considered but they either vanish or are very
small and we discuss them briefly in the appendices.

\section{$N^*(1520)-h$ contribution to the $\rho $ meson selfenergy}
Next we intend to modify our formalism so that the $\rho$ meson is able
to excite baryonic resonances via a $\rho N N^*$ coupling. From now on we will
restrict the discussion to the case of $N^* (1520)$ $I(J^P)=\frac{1}{2}
(\frac{3}{2} ^{-})$, although the same arguments considered here may be used to
introduce other resonances with different quantum numbers \footnote{Note that
since we stay in the center of mass frame of the $\rho$ meson, the contribution
of resonances which couple to the $\rho$ in P-wave vanishes. The coupling to
$N^* (1520)$ is S-wave and therefore its contribution to the $\rho$ selfenergy
survives even at zero $\rho$ three-momentum.}. 

\begin{figure}[ht]
\centerline{\includegraphics[width=0.7\textwidth]{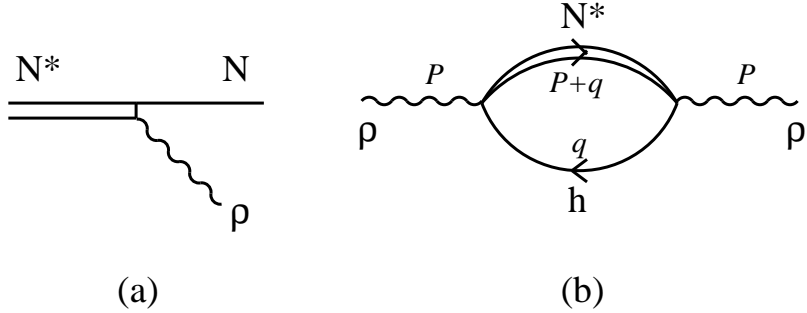}}
\caption{\footnotesize{(a) $\rho N N^*$ vertex. (b) $N^*-h$ bubble contributing
to the selfenergy of the $\rho$ meson.}}
\label{diagNstar}
\end{figure}

The correction under discussion (excitation of a $N^* -h$ pair) will manifest as
an extra selfenergy term in the $\rho$ propagator.
The basic vertex involved in this effect is shown in Fig.
\ref{diagNstar}a, and the lagrangian describing the
interaction reads \cite{osetcanogomez}
\ba
\label{LN*Nrho}
\mathcal{L}_{N^* N \rho} = -g_{N^* N \rho} \bar{\Psi}_N S_i \vec{\phi_i} \vec{\tau} \Psi_{N^*} + h.c., 
\ea
where $\Psi_N$, $\Psi_{N^*}$, $\phi_i$ are the $N$, $N^*(1520)$, $\rho$ fields, $S_i$ is the $\frac{1}{2} \to
\frac{3}{2}$ spin transition operator, $\vec{\tau}$ is the isospin-$\frac{1}{2}$
operator and $g_{N^* N \rho}$ stands for 
the $\rho N N^*$ coupling constant that we take from
\cite{osetcanogomez} to be $g_{N^* N \rho} = 7.73/\sqrt{3}$.
Feynman rules from the lagrangian in eq. (\ref{LN*Nrho}) allow us to calculate
the selfenergy corresponding to the graph \ref{diagNstar}b
and we get
\ba
\label{PINstar1}
-i \Pi_{\rho}^{N^*-h} (P) = - \int{\frac{d^{4}q}{(2\pi)^{4}} (IF) (SF) g_{N^* N \rho}^2
\frac{1}{P^0+q^0-\mathcal{E}_{N^*} (\vec{P}+\vec{q}) + i\epsilon}
\frac{n(\vec{q})}{q^0-\mathcal{E}_{N} (\vec{q}) - i\epsilon}}.
\ea
In eq. (\ref{PINstar1}), $(IF)=2$ and $(SF)=\frac{4}{3}$ are isospin and spin
factors respectively, $n(\vec{q})$ stands for the
occupation number of the nucleonic hole and $\mathcal{E}_B(\vec{q})$ is the energy of the baryon $B$ involved,
\ba
\label{E_B}
\mathcal{E}_B(\vec{q}) = \sqrt{M_B^2+\vec{q}\,^2} \simeq M_B + \frac{\vec{q}\,^2}{2 M_B}.
\ea
The $q^0$ integration can be performed analytically, and we are left with
\ba
\label{PINstar2}
\Pi_{\rho}^{N^*-h} (P) &=& \frac{2}{3} g_{N^* N \rho}^2 U_{N^*} (P) \nonumber \\
U_{N^*} (P) &=& 4 \int{\frac{d^{3}q}{(2\pi)^{3}} \frac{n(\vec{q})}{P^0 +
\mathcal{E}_{N} (\vec{q}) - \mathcal{E}_{N^*} (\vec{P}+\vec{q})}},
\ea
where we have defined the Lindhard function for the $N^*-h$ excitation, 
$U_{N^*} (P)$. In both eqs. (\ref{PINstar1}) and (\ref{PINstar2}) we have
omitted the decay width in the $N^*$ propagator, although it is included in the
calculation (see Appendix).

Following the structure of the $\rho$ meson propagator from the $T$ matrix discussed in
the previous section, now it is easy
to include a selfenergy term coming from a $\rho N N^*$ coupling, just adding it to the
one already existing from the coupling of the $\rho$ meson to $\pi \pi$ pairs,
nucleons and deltas:  
\begin{eqnarray}
\label{rhoself}
\Pi \to \Pi  + \Pi_{\rho}^{N^*-h}. 
\end{eqnarray}

\section{Results and discussion}

\begin{figure}[ht]
\centerline{\includegraphics[width=0.7\textwidth]{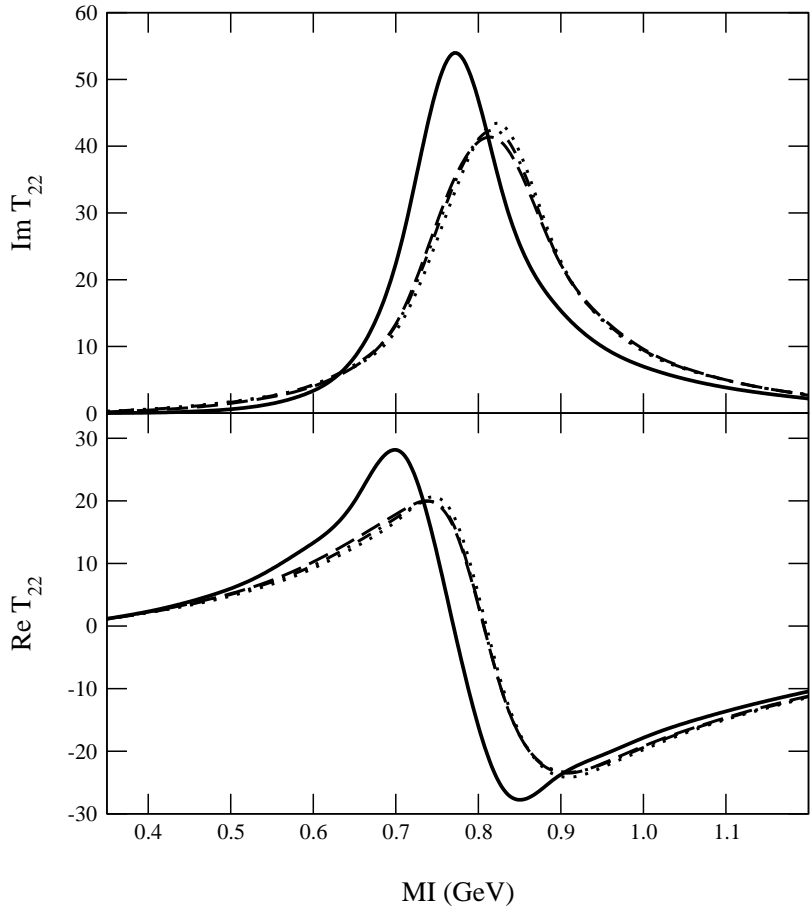}}
\caption{\footnotesize{Real and imaginary parts of the $\pi\pi \to \pi\pi$
scattering amplitude vs invariant mass. Solid lines stand for the free case. The
others correspond to the calculation at $\rho=\rho_0$: dashed-dotted lines for
$\Lambda=1$ GeV, long dashed and dotted lines for $\Lambda=0.9,1.1$ GeV
respectively.}} 
\label{cambio_L}
\end{figure}

First of all we show the dependence of our results on parameters that enter the
calculation of the pion selfenergy, one of the essential ingredient in the
renormalization of the $\rho$ up to this point.
In Fig. \ref{cambio_L} we show the $\pi \pi \to \pi \pi$ scattering amplitude
vs invariant mass. The calculation has been performed neglecting the $N^*-h$
excitation. The $g'$
parameter in the pion selfenergy is set to 0.7. In the figure we plot the real
and imaginary parts for the free case and at normal nuclear matter density.
Compared to the free case, the imaginary part shows a clear 
broadening which is expected since there are new decay channels. At the
same time one can observe a shift of the peak of the $\rho$ distribution to
higher energies. The resonance shape of the distribution
remains at $\rho=\rho_0$ and the zero of the real part of the
amplitude also moves to higher energies.
At normal nuclear density, the shift amounts to 
about $30-40$ MeV and the width at half maximum is somewhat more than 200 MeV to
be compared to the free value of 150 MeV. We have changed the $\pi N N$, $\pi N
\Delta$ form factor
parameter $\Lambda$ between $900$ MeV and $1100$ MeV. The differences found are
small and this can give an 
idea of the uncertainties in the present results which we can expect from
uncertainties in the pion selfenergy. Other uncertainties coming from 
details on the treatment of the $ \Delta$ resonance will be discussed later.
\vspace{0.5cm}

\begin{figure}[ht]
\centerline{\includegraphics[width=0.7\textwidth]{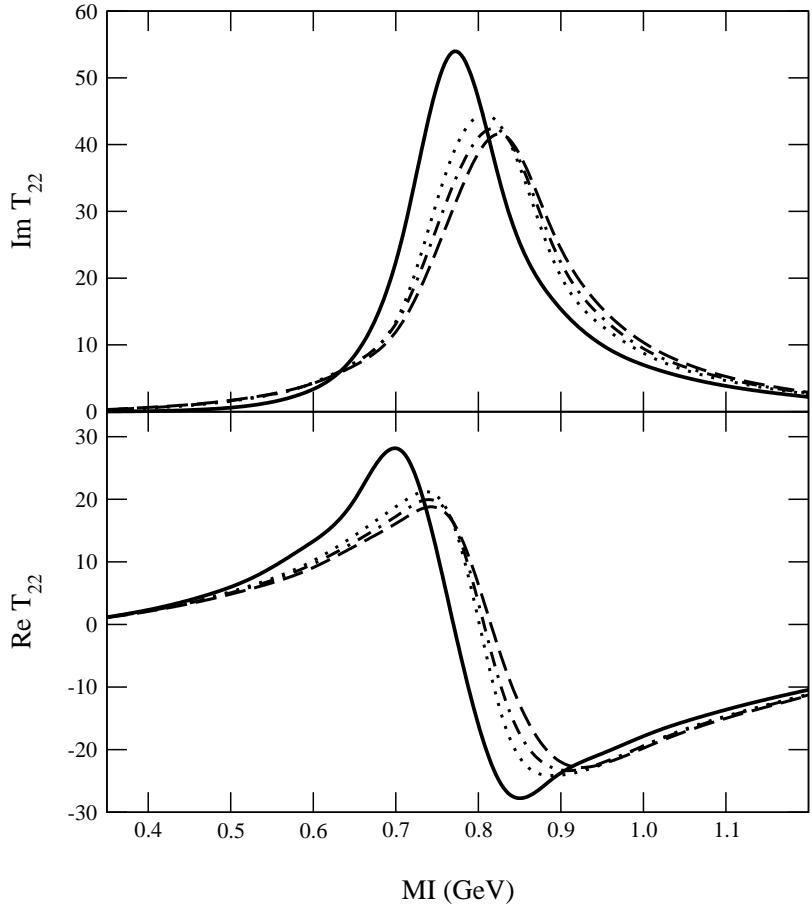}}
\caption{\footnotesize{The same as in Fig. \ref{cambio_L} for $\rho=\rho_0$
and several values of $g'$. Dashed-doted lines stand for $g'=0.7$, long dashed
and dotted lines being for $g'=0.6,0.8$ respectively.}}
\label{cambio_g}
\end{figure}

We have also checked the dependence of the calculation on the $g'$ Landau-Migdal
parameter appearing in eq. (\ref{Piself1}). The results of this check are
presented in Fig. \ref{cambio_g}, in which we show the real and imaginary parts
of the scattering amplitude at $\rho=\rho_0$. The free case is also displayed
for reference. The effects of a
varying $g'$ are noticeable in the resonance region whereas the results are
insensitive for lower and higher energies.
Neither the position of the
resonance nor its width are much changed with variations of $g'$ in the
standard range of values $[0.6,0.8]$. We find a fluctuation of the peak
position of around 3 \% of the total mass at $\rho=\rho_0$.
\vspace{0.5cm}

The results on the mass shift presented so far are in qualitative agreement with
other works which find a moderate shift of the $\rho$ meson mass to higher
energies at finite densities by using different approaches
\cite{Chanfray:1993ue,Urban:1998eg}.

The treatment of the $\Delta (1232)$ resonance is another issue with remarkable
impact on the spectral behaviour of the $\rho$ meson in nuclear matter. 
The works of refs. \cite{Chanfray:1993ue,Urban:1998eg} where the main
ingredient is the medium pion selfenergy mostly driven by the coupling to
$\Delta -h$ excitations find also a broadening of the $\rho$ meson. We
have found 
that the shape of the spectral distribution at low energies depends
significantly on the $\Delta$ resonance 
properties specially its decay width. For instance, a simplified treatment in
which a constant width is assumed gives a
bump at around 500 MeV for $\rho \gtrsim \rho_0$. This bump disappears when an
energy dependent width is 
used (see also \cite{Rapp:2000ej}, Fig. 3.13). Nonetheless, in some calculations
having explicit energy dependence on the $\Delta$ width the bump appearing at
low energies in the spectral function of the $\rho$ meson, although weakened,
still remains.

\begin{figure}[ht]
\centerline{\includegraphics[width=0.7\textwidth]{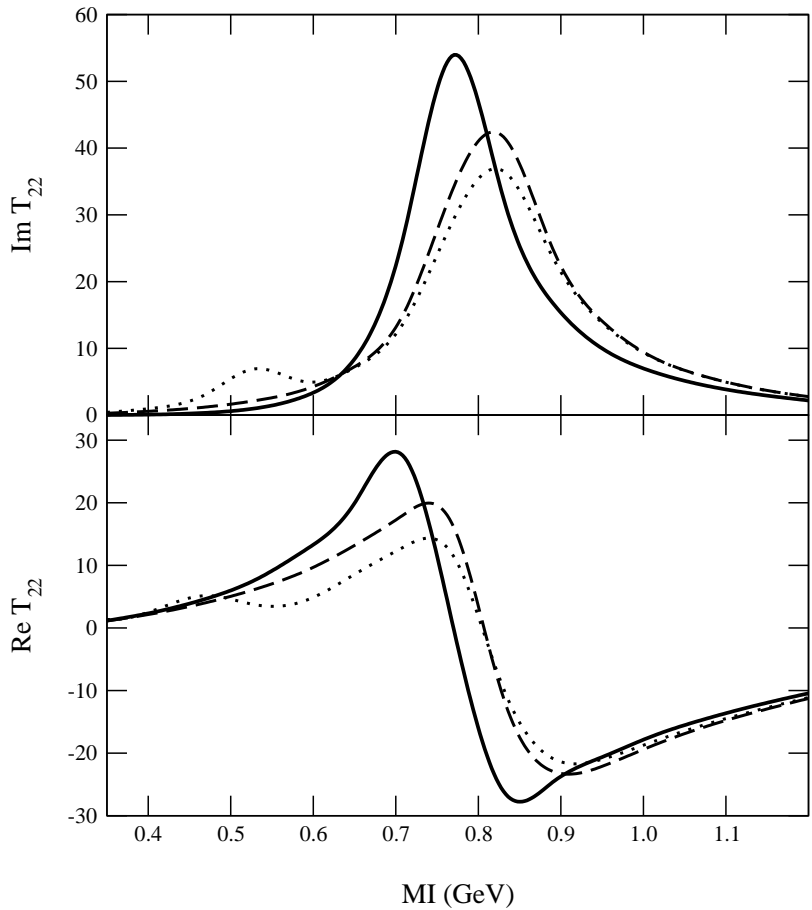}}
\caption{\footnotesize{Real and imaginary parts of the $\pi\pi \to \pi\pi$
scattering amplitude. Long dashed (dotted)  lines
correspond to the results without (with) the effects of the coupling $\rho N
N^*$ at $\rho=\rho_0$, and solid lines stand for the free case.}}   
\label{meto_Nstar}
\end{figure}

Next we include in the calculation the $N^*(1520)$ contribution to the $\rho$
meson selfenergy in nuclear matter, as described in section 4. 
Our model does not try to be complete since many other resonances should be
included. A much detailed work along these lines can be found in ref. 
\cite{Peters:1998va}. Our aim is simply to estimate the effect of these type of
decay channels on our previous results.
The $\pi \pi$ amplitude
is plotted in Fig. \ref{meto_Nstar} for $\rho=\rho_0$ with and without the $N^*
-h$ contribution, together with the free
case for reference. The possibility of exciting a $N^*-h$ pair
is reflected in the structure arising in the amplitude around $550$ MeV. As a
consequence a sizeable amount of strength appears at energies below the
$\rho$ meson mass.
Our results agree qualitatively with other works in which the $\rho$ meson
is allowed to couple to baryonic resonances \cite{Peters:1998va,Lutz:1999jn}.
\vspace{0.5cm}

\begin{figure}[ht]
\centerline{\includegraphics[width=0.7\textwidth]{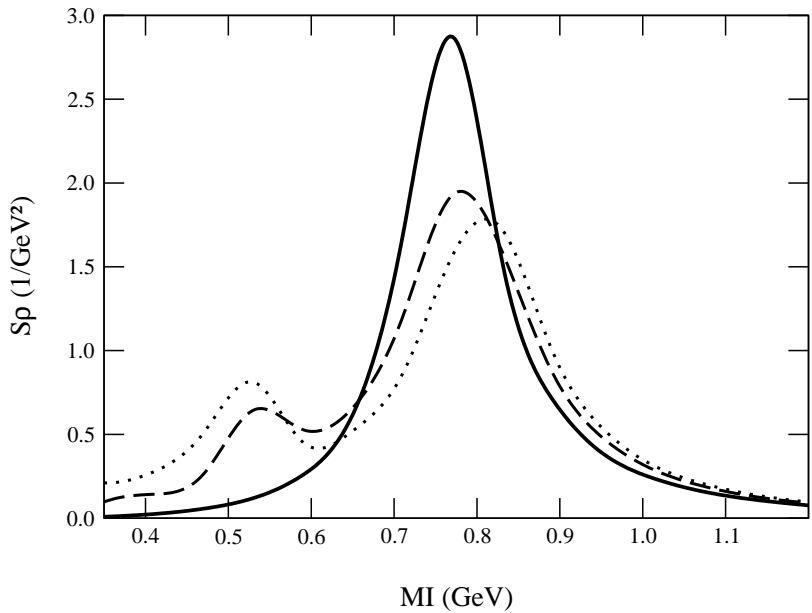}}
\caption{\footnotesize{Spectral function of the $\rho$ meson for several values
of nuclear matter density. Lines are as follows: solid, dashed and dotted stand
respectively for $\rho=0$, $\rho_0 /2$ and $\rho_0$.}}
\label{cambiodensidad}
\end{figure}

The same behaviour discussed above is reflected in the in-medium spectral
function $S_{\rho}=-\frac{1}{\pi} $Im$ D_{\rho}$, which we show in Fig.
\ref{cambiodensidad} for the free case and several values of nuclear matter
density. 
The first effect to be commented is that the peak of the $\rho$ distribution
broadens substantially as the density is increased,
its width being about 200 MeV for $\rho=\rho_0$. This is accompanied by a
clear shift of the position of the resonance to higher energies (in agreement
with many other works, see \cite{Rapp:2000ej} for reference) that amounts, as
reported before, 
to about $30-40$ MeV for normal nuclear matter density. Together with the $\rho$
peak we can observe a resonant structure corresponding to the excitation of the
$N^*(1520)$ appearing around 500-550 MeV. This increases noticeably the amount
of strength towards low energies, which goes in the right direction in what
concerns the description of dilepton radiation spectra in that energy region
\cite{Rapp:2000ej}. The strength of the $N^*(1520)$ peak increases with
density, which is expected since this is an effect linear
in density in our model. It is also noticeable that the position of this
structure is slightly pushed downwards in energy with increasing density, as
a consequence of the interference with the other contributions to the $\rho$
selfenergy. In this respect we agree with refs. \cite{Friman:1998fb,Lutz:1999jn}
where it was pointed out that because of the coupling to baryon resonances, the
in medium vector meson strength is split into a meson like mode, which is pushed
up in energy and a resonance-hole like mode, which is pushed down in energy.

The results would change if one takes into account that the properties of
baryonic resonances depend themselves on the rho meson medium spectral function,
as it is indeed the case for the $N^* (1520)$ since it decays, for instance,
into $N \rho [\pi \pi]$. Therefore a selfconsistent treatment is advisable.
This has been done in \cite{Peters:1998va}, where a melting of both
$N^*$ and $\rho$ meson structures is found at high densities so that none of
them are distinguishable any longer. As a consequence, much strength is spread
to lower invariant masses. Note however that this result cannot be extrapolated
directly to our case since we obtain a weaker contribution of the $N^* -h$
excitation and our in-medium $\rho$ is also narrower than in
\cite{Peters:1998va}. 

Our results reflect the most important characteristics of the medium corrections
and these are shared qualitatively by many other works in which different
approaches to the problem have been used
\cite{Klingl:1997kf,Chanfray:1993ue,Herrmann:1993za,Urban:1998eg,Peters:1998va,Lutz:1999jn}.

\section{Summary}
We have analyzed the problem of the $\rho$ properties in a nuclear medium from
a new perspective, using a chiral unitary framework to describe the $\rho$ meson
in free space and in the medium.

The approach starts from the lowest order chiral lagrangian involving the
pseudoscalar mesons plus those lagrangians coupling vector mesons to
pseudoscalar mesons. 
The unitarization for meson-meson scattering is done following closely the idea
of the $N/D$ method in coupled channels and leads to an iteration of the tree
level
diagrams obtained from the chiral lagrangians plus the exchange of bare vector
mesons. The free parameters of the theory are obtained by matching the results
to those of chiral perturbation theory for the $\pi$ and $K$ form factors plus
the position of the $\rho$ peak. Once this is done the theory reproduces the
experimental form factors and the scattering amplitudes in the vector sector
with high accuracy up to $1.2$ GeV.

The treatment of the $\rho$ in the medium has led us to modify the previous
approach, which relies upon the on-shell amplitudes provided by the lagrangians
and dimensional regularization, in order to properly incorporate pion
selfenergies and vertex corrections in a gauge invariant way. Whereas in vacuum
it has been found that a formalism keeping tadpole diagrams and full off-shell
dependence of the $\rho \pi \pi$ vertex is equivalent to the one followed in
ref. \cite{Oller:2001ug}
where only the on shell part of the $\rho \pi \pi$ vertex is kept and no tadpole
diagrams are considered, in the medium one should explicitly include tadpole
terms in order to preserve gauge invariance.

The inclusion of medium effects proceeds by dressing the
pion propagators with a proper selfenergy and including vertex corrections. In
this way the main $\rho$ decay channel into $\pi \pi$ is modified and new decay
channels are opened. The main visible effect of the nuclear medium is a
broadening of the $\rho$ width as the nuclear density increases. Simultaneously
we observe that there is a small shift of the peak to higher energies
moderately dependent on the uncertainties of the pion selfenergy.

We have also added another source of medium $\rho$ renormalization including the
coupling of the $\rho$ to the $N^*(1520) -h$ component. This part is not novel
and has been studied elsewhere, but we include it for completeness in order to
show how our previous results are affected by this channel. We observe that an
important source of strength appears in the $\rho$ spectral function at low
energies and that the previous peak of the $\rho$ distribution is slightly
pushed to higher energies. Yet the global effect is still a shift of the $\rho$ spectral
function peak to higher energies of about $30-40$ MeV at $\rho=\rho_0$.
Our results would thus agree with those where there is not much
shift of the $\rho$ peak but however produce an extra strength in the
isovector-vector channel at lower energies than the free $\rho$ mass, although
there are quantitative differences.

\bigskip

\subsection*{Acknowledgments}
We acknowledge partial financial support from the DGICYT under
contract BFM2000-1326 and from the EU TMR network Eurodaphne, contract no. 
ERBFMRX-CT98-0169. D.~C. thanks financial support from MCYT.

\section*{Appendix: Formulae for the $N^*-h$ Lindhard function}
We give the analytic expression for the Lindhard function \cite{ReportdeEulogio}
corresponding to the $N^*-h$ excitation, quoted in eq. (\ref{PINstar2}):
\ba
\label{NstarLind}
U_{N^*} (P^0,\vec{P};\rho) = \frac{3}{2} \rho \frac{M_{N^*}}{P k_{F}} 
\bigg \lbrack z + \frac{1}{2} (1-z^2) \log \frac{z+1}{z-1}
+ z_c + \frac{1}{2} (1-z_c^2) \log \frac{z_c+1}{z_c-1} \bigg \rbrack,
\ea
where $k_F$ is the Fermi momentum, and here $P$ stands
for the modulus of $\vec{P}$. Terms depending on $z$, $z_c$ come from direct and
crossed diagrams, respectively. These functions are defined as
\ba
\label{z}
z &=& \frac{M_{N^*}}{P k_{F}} \bigg \lbrack P^0 - \omega_R - 
\frac{P^2}{2 M_{N^*}} 
+ \frac{i}{2} \Gamma(P^0,\vec{P}) \bigg \rbrack \nonumber \\
z_c &=& \frac{M_{N^*}}{P k_{F}} \bigg \lbrack -P^0 - \omega_R - 
\frac{P^2}{2 M_{N^*}} 
+ \frac{i}{2} \Gamma(-P^0,\vec{P}) \bigg \rbrack,
\ea
with $\omega_R = M_{N^*}-M_N$. Eq. (\ref{z}) already includes the information of
the $N^*$ decay width, which can be found in \cite{osetcanogomez}.

The limit of eq. (\ref{NstarLind}) when $\vec{P} \to \vec{0}$ is given by
\ba
\label{Limit}
U_{N^*} (P^0,\vec{0};\rho) = \rho \bigg \lbrack \frac{1}{P^0 - \omega_R + 
\frac{i}{2} \Gamma(P^0,\vec{0})} + \frac{1}{-P^0 - \omega_R} \bigg \rbrack,
\ea
and it is the expression that we use in our calculation as we stay in the CM
frame of the $\rho$ meson.

\section*{Appendix 2: Other medium corrections not included in the text}
\begin{figure}[ht]
\centerline{\includegraphics[width=1.0\textwidth]{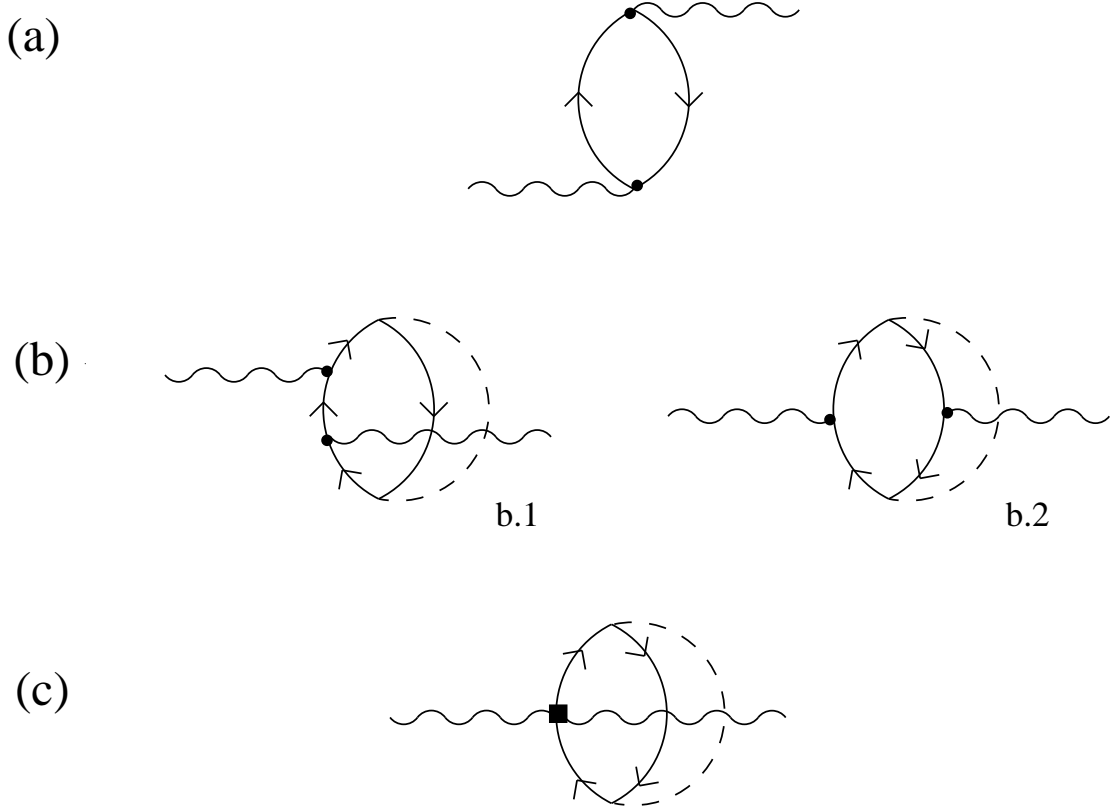}}
\caption{\footnotesize{Medium correction graphs not included in the calculation:
sets (a) and (b) involving zero and one pion lines. $\rho N N$, $\rho \Delta
\Delta$ vertices are displayed as dots.}}
\label{others}
\end{figure}

In section 3 we have described how the presence of nuclear matter modifies the
propagation of the $\rho$ meson by modifying the properties of the pionic cloud.
In addition to the pion selfenergy, a set of diagrams involving the $\rho \pi N
N$ contact term were also considered to preserve gauge invariance (see Fig.
\ref{diaggauge}). Minimal substitution, which provides the interaction
lagrangians for the mesons and baryons involved
\cite{Herrmann:1993za,Urban:1998eg}, gives rise to some other diagrams that we
have not included in the text because (i) they are higher powers of density (in
the calculation only corrections of order $\rho$ are considered), (ii) they
vanish for vanishing $\rho$
momentum or because of symmetry reasons, or (iii) they are negligible compared
to other corrections. Now 
we focus on some of them and outline the derivation of some expressions that
lead to our conclusions. For this purpose we classify the graphs in two
sets, depending on the number of pion propagators (zero or one pionic
lines), as it is shown in Fig. \ref{others}.

\subsection*{\emph {No pion lines}}
The first vanishing contribution corresponds to the diagram labelled as
\ref{others}a, in which a $\rho$ meson directly couples to a $p-h$ excitation.
This mechanism does not contribute for a $\rho$ meson at rest, since then both
the particle and the hole have the same momentum. The similar diagram with
$\Delta -h$ excitation involves a vanishing vertex in the non relativistic
approximation, for a $\rho$ at rest.

\subsection*{\emph {One pion line}}
Here we distribute the graphs in three subsets: both $\rho$ mesons coupled to
either a particle or a hole line (Fig. \ref{others}b.1), one
$\rho$ meson line attached to the particle line and the other one to the hole
line (Fig. \ref{others}b.2) and both $\rho$ mesons involved in a $\rho\rho N N$
contact vertex (Fig. \ref{others}c).
\begin{figure}[ht]
\centerline{\includegraphics[width=0.7\textwidth]{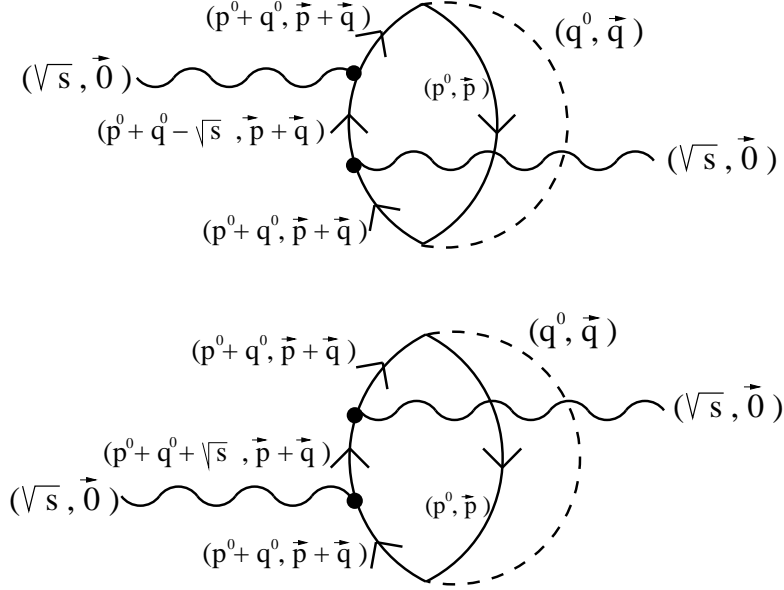}}
\caption{\footnotesize{Momentum labelling of the diagram with two $\rho$ mesons
coupled to the particle line.}}
\label{b1detail}
\end{figure}

Let us calculate the contribution of diagrams \ref{others}b.1. to the $\rho$
meson selfenergy. We focus on the case in which both incoming and outgoing
$\rho$ mesons are coupled to the particle line (in Fig. \ref{b1detail} the
momentum labelling is shown). Applying the Feynman rules derived from the interaction
lagrangians of \cite{Urban:1998eg} we get the following expression:

\begin{eqnarray}
\label{b1}
- i \Pi^{ij}_{(b.1)} & = & - i \delta^{ij} K \int \frac{d^{4}q}{(2\pi)^{4}}
(\vec{q}\,^2)^2 D(q) \int \frac{d^{4}p}{(2\pi)^{4}} \bigg \lbrace G_h(p)
G_p^2(p+q) G_p(p^0+q^0+\sqrt{s},\vec{p}+\vec{q}) +{} \nonumber \\
& &{}G_h(p) G_p^2(p+q) G_p(p^0+q^0-\sqrt{s},\vec{p}+\vec{q}) \bigg \rbrace
\end{eqnarray}
where $G_p$ ($G_h$) stands for the propagator of a particle (hole) and $K = 
(\frac{f_N}{m_{\pi}})^2 g_{\rho}^2 (\frac{1}{M_N})^2$. In obtaining the previous
expression, the $\rho N N$ vertices 
provide a $(\vec{p}+\vec{q})_i (\vec{p}+\vec{q})_j$ factor. The terms linear and
quadratic in $\vec{p}$, when performing the integration over the momenta in the
fermionic loop give rise to contributions of order $\rho^{\frac{4}{3}}$ and
higher powers in the nuclear density. We neglect those contributions and then
eq. (\ref{b1}) follows. 
This result can be expressed in terms of the ordinary Lindhard functions. The
product of fermion propagators in eq. (\ref{b1}) can be rewritten as

\begin{eqnarray}
\label{trick}
\lefteqn{G_h(p) G_p^2(p+q) G_p(p^0+q^0+\sqrt{s},\vec{p}+\vec{q}) ={} }
\nonumber \\
& & {}\frac{\partial}{\partial \alpha} \bigg \lbrace \frac{1}{\sqrt{s}+\alpha}
\bigg \lbrack G_h(p) G_p(p^0+q^0-\alpha,\vec{p}+\vec{q}) - G_h(p)
G_p(p^0+q^0+\sqrt{s},\vec{p}+\vec{q}) \bigg \rbrack \bigg \rbrace \nonumber \\
\lefteqn{G_h(p) G_p^2(p+q) G_p(p^0+q^0-\sqrt{s},\vec{p}+\vec{q}) ={} }
\nonumber \\
& & {}\frac{\partial}{\partial \alpha} \bigg \lbrace \frac{-1}{\sqrt{s}-\alpha}
\bigg \lbrack G_h(p) G_p(p^0+q^0-\alpha,\vec{p}+\vec{q}) - G_h(p)
G_p(p^0+q^0-\sqrt{s},\vec{p}+\vec{q}) \bigg \rbrack \bigg \rbrace
\end{eqnarray}
where the limit $\alpha \to 0$ is assumed. Commutation of this limit and the
$\partial_\alpha$ derivative with the integration over the fermionic loop
momenta allows us to write

\begin{eqnarray}
\label{b1II}
- i \Pi^{ij}_{(b.1)} = \delta^{ij} \frac{K}{2} \int \frac{d^{4}q}{(2\pi)^{4}} (\vec{q}\,^2)^2 
D(q) \frac{1}{s} \bigg \lbrack \bar{U}_N (q^0+\sqrt{s},\vec{q}) + \bar{U}_N 
(q^0-\sqrt{s},\vec{q}) - 2 \bar{U}_N (q) \bigg \rbrack
\end{eqnarray}
where the $\bar{U}$ function, the Lindhard function for a forward propagating
$p-h$ bubble, stands for
\begin{eqnarray}
\label{Ubarra}
\bar{U}_N (q) = -2i \int \frac{d^{4}p}{(2\pi)^{4}} G_h(p) G_p(p+q)
\end{eqnarray}
and satisfies
\begin{eqnarray}
\label{UrelatesL}
2 \, \lbrack \bar{U}_N (q^0,\vec{q}) + \bar{U}_N (-q^0,\vec{q}) \rbrack = U_N
(q^0,\vec{q}).
\end{eqnarray}
We have compared numerically the size of this contribution to any of the ones included in
Fig. \ref{diagmed} and we find that it is around a hundred times smaller in the
range of invariant mass under study. Including this contribution in the
calculation has no visible effect and therefore we neglect it, and so we do with
the terms containing $\Delta$ baryons instead of nucleons.

In this derivation we have neglected all terms linear or quadratic in the hole
momentum, since they generate contributions of higher order in density.
Following this criterium, we can discard other contributions from diagrams with
a $\rho$ meson coupled to a hole line. In particular this is the case of the
partners of the graphs shown of Fig. \ref{b1detail}, with both $\rho$ mesons
coupled to the hole line. The same happens to the diagrams of the type of the
one shown in Fig. \ref{others}b.2.

We are left with the evaluation of the $\rho \rho \pi \pi$ vertex correction
shown in Fig. \ref{others}c, where although not displayed, the $\rho\rho N N$
vertex can also be inserted in the hole line. The momentum labelling follows the
one in Fig. \ref{b1detail}. Using the Feynman rule for the $\rho\rho N N$ vertex
derived in section 3.6.3 we have
\begin{eqnarray}
\label{rhoNrhoNpionselfenergy}
-i \Pi^{ij} = -i g_{\rho}^2 \frac{3}{M_N} \bigg ( \frac{f_N}{m_{\pi}} \bigg ) ^2
\delta ^{ij} 
\int \frac{d^4 q}{(2\pi)^4} \vec{q}\,^2 D(q) \int \frac{d^4 p}{(2\pi)^4} G(p)
G(p+q)^2,
\end{eqnarray}
where here $G(p)$ denotes the whole nucleon propagator with both the particle
and the hole pieces. It is possible to rewrite $G(p+q)^2$ as
\begin{equation}
\label{trickkk}
G(p+q)^2 = \partial_{\alpha} G(p^0+q^0-\alpha,\vec{p}+\vec{q}),
\end{equation}
with $\alpha \to 0$. The integration of the fermionic loop is then given by
\begin{eqnarray}
\label{fermloop}
I =  \int \frac{d^4 p}{(2\pi)^4} G(p) G(p+q)^2 = \partial_{\alpha} \int
\frac{d^4 p}{(2\pi)^4} G(p) G(p^0+q^0-\alpha,\vec{p}+\vec{q}) =
\nonumber \\
\frac{i}{4} \partial_{\alpha} U(q^0-\alpha,\vec{q}) |_{\alpha=0} ,
\end{eqnarray}
which is an odd function of $q^0$ since the Lindhard function is even in $q^0$.
As a consequence the $dq^0$ integration vanishes and the contribution of these
terms is null. Similar arguments for the case of a $\Delta$ in the particle line
also lead to a null contribution.

\end{document}